\documentclass{article}

\usepackage{url}
\usepackage{graphicx}
\usepackage{wrapfig}
\usepackage{caption}
\usepackage{enumitem}
\usepackage{amsbsy}
\usepackage{bbm}
\usepackage[utf8]{inputenc} 
\usepackage[T1]{fontenc}    
\usepackage{hyperref}       
\usepackage{url}            
\usepackage{booktabs}       
\usepackage{amsfonts}
\usepackage[table,xcdraw]{xcolor}
\usepackage{nicefrac}       
\usepackage{microtype}      
\usepackage{xcolor}         
\usepackage{subfigure}
\usepackage{amsmath}
\usepackage{amssymb}
\usepackage{booktabs}
\usepackage{adjustbox}
\usepackage{multirow}
\usepackage{caption}
\usepackage{booktabs}
\usepackage{wrapfig}
\usepackage{dsfont}
\usepackage{titletoc}
\usepackage{lipsum}
\usepackage{subcaption}
\usepackage{float} 
\usepackage{algpseudocode}
\usepackage[linesnumbered,ruled,vlined]{algorithm2e}
\usepackage{amsmath}
    
\usepackage{amssymb}
  \makeatletter
  
  \newcommand{\Rmnum}[1]{\expandafter\@slowromancap\romannumeral #1@}
  \makeatother
\usepackage[preprint]{neurips_2025}
\usepackage{enumitem}
\usepackage{amsbsy}
\usepackage{natbib}  
\usepackage{bbm}
\usepackage[utf8]{inputenc} 
\usepackage[T1]{fontenc}    
\usepackage{hyperref}       
\usepackage{url}            
\usepackage{booktabs}       
\usepackage{amsfonts}       
\usepackage{nicefrac}       
\usepackage{microtype}      
\usepackage{xcolor}         
\usepackage{subfigure}
\usepackage{amsmath}
\usepackage{amssymb}
\usepackage{graphicx}
\usepackage{booktabs}
\usepackage{adjustbox}
\usepackage{multirow}
\usepackage{caption}
\usepackage{booktabs}
\usepackage{wrapfig}
\usepackage{lipsum}
\usepackage{float}
\usepackage{algpseudocode}
\usepackage[linesnumbered,ruled,vlined]{algorithm2e}
\usepackage{amsmath}
\makeatother
\title{Fine-Grained Privacy Extraction from Retrieval-Augmented Generation Systems via Knowledge Asymmetry Exploitation}

\author{%
  Yufei Chen\\
  Xidian University \\
  \texttt{24151111267@stu.xidian.edu.cn}
  \And
  Yao Wang\thanks{Corresponding author.}\\
  Xidian University \\
  \texttt{wangyao@xidian.edu.cn}
  \And
  Tao Gu\\
  Macquarie University \\
  \texttt{tao.gu@mq.edu.au}
  \And
  Haibin Zhang\\
  Xidian University \\
  \texttt{hbzhang@mail.xidian.edu.cn}
}

\begin{document}

\maketitle

\begin{abstract}
Retrieval-Augmented Generation (RAG) systems enhance large language models (LLMs) by incorporating external knowledge bases, significantly improving their factual accuracy and contextual relevance. However, this integration also introduces new privacy vulnerabilities. Existing privacy attacks on RAG systems may trigger data leakage, but they often fail to accurately isolate knowledge base-derived content within mixed responses and perform poorly in multi-domain settings. In this paper, we propose a novel black-box attack framework that exploits knowledge asymmetry between RAG systems and standard LLMs to enable fine-grained privacy extraction across heterogeneous knowledge domains. Our approach decomposes adversarial queries to maximize information divergence between the models, then applies semantic relationship scoring to resolve lexical and syntactic ambiguities. These features are used to train a neural classifier capable of precisely identifying response segments that contain private or sensitive information. Unlike prior methods, our framework generalizes to unseen domains through iterative refinement without requiring prior knowledge of the corpus. Experimental results show that our method achieves over 90\% extraction accuracy in single-domain scenarios and 80\% in multi-domain settings, outperforming baselines by over 30\% in key evaluation metrics. These results represent the first systematic solution for fine-grained privacy localization in RAG systems, exposing critical security vulnerabilities and paving the way for stronger, more resilient defenses.

\end{abstract}

\section{Introduction}

LLMs have revolutionized natural language processing \citep{brown2020language}, yet they face critical limitations in domain-specific tasks, such as hallucinations and outdated information \citep{kandpal2023large, shuster2021retrieval, gao2023retrieval}. RAG addresses these issues by integrating external knowledge bases \citep{lewis2020retrieval}, enabling more accurate and context-aware responses in domains such as medical \citep{xiong2024benchmarking}, financial \citep{yepes2024financial}, legal consulting \citep{mahari2021autolaw}, and personal assistant applications \citep{liu2020towards}. However, this architectural advantage introduces a significant risk: when knowledge bases contain sensitive data (e.g., medical records and financial documents), RAG systems may inadvertently expose private information through their outputs.

RAG systems face two main types of privacy-related attacks. Membership inference attacks analyze response patterns to determine if a document exists in the knowledge base, using techniques such as document fragment masking \citep{liu2024mask} or prediction score calculation \citep{anderson2024my, shi2023detecting}. However, these attacks require exact copies of target documents, an impractical requirement for private knowledge bases where data is typically unique or obfuscated. Privacy extraction attacks, meanwhile, use carefully crafted prompts to induce RAG systems to leak private data, but suffer from two fundamental limitations.

First, they achieve only \textbf{coarse-grained} leakage detection, capable of determining that private data is present in responses but unable to identify which specific sentences originate from the knowledge base \citep{qi2024follow, zeng2024good}. This is because RAG responses blend external knowledge with the LLM’s pre-trained content, creating an information mixture problem that confounds source determination. While regular expressions can extract private data, this method only works with fixed data structures \citep{jiang2024rag}. The inherent diversity and randomness of LLM-generated text, which lacks unified structural features, makes accurate privacy identification challenging.

Second, existing methods are limited to \textbf{single-domain} scenarios with concentrated knowledge and coherent contexts \citep{qi2024follow,zeng2024good,jiang2024rag}. They struggle to adapt to real-world RAG systems that handle diverse domains (e.g., insurance platforms combining health records, policy details, and claim regulations). In cross-domain data, the scattered knowledge distribution and wide topic range make it difficult to construct targeted adversarial queries, significantly reducing attack effectiveness.

\begin{wrapfigure}{r}{0.55\textwidth}
\centering
\vspace{-20pt}
\includegraphics[width=0.55\textwidth]{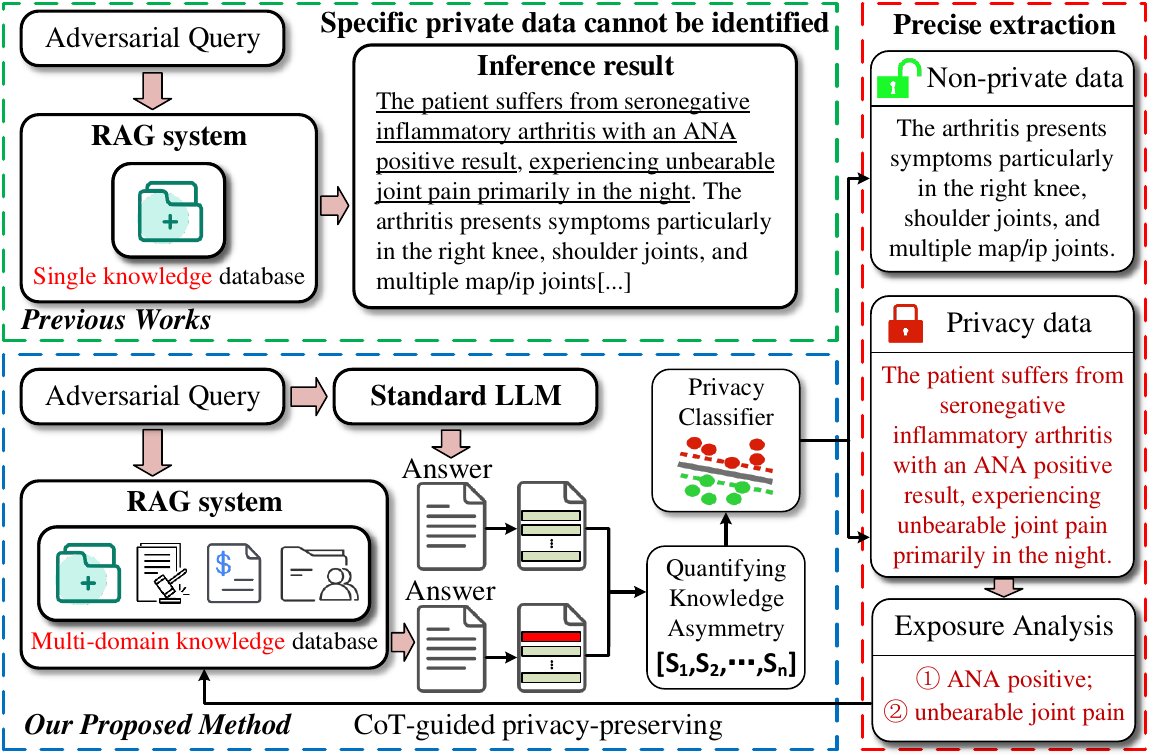}
\captionsetup{labelfont=bf}
\caption{Comparison of our attack with existing work.}
\label{Introduction}
\vspace{-10pt} 
\end{wrapfigure}

To address these challenges, we introduce a novel black-box attack framework that achieves \textbf{fine-grained} private data localization in both single and \textbf{multi-domain} RAG systems. As illustrated in Figure \ref{Introduction}, previous works fail to distinguish private from non-private contents in inference results when processing adversarial queries. Our approach precisely identifies both types of data through quantitative analysis of the knowledge asymmetry between standard LLMs and RAG systems. RAG systems incorporate external knowledge absent from standard LLMs' pre-trained knowledge, creating inherent semantic discrepancies. These systematic differences provide critical cues for accurately extracting 
private information.

Our framework first generates adversarial queries designed to elicit detailed RAG responses while amplifying differences from standard LLMs. To address the challenges of scattered and heterogeneous knowledge in multi-domain environments, we introduce an iterative query refinement strategy. This approach uses broad, domain-agnostic initial queries to detect potential privacy leakage by observing response divergence. It then iteratively feeds extracted privacy fragments back to refine subsequent queries. This mechanism enables our method to synthesize targeted adversarial queries in zero-prior-knowledge scenarios, ensuring the RAG system fully leverages its knowledge base. We then segment responses into sentences, convert them to semantic vectors, and measure their similarity. Since cosine similarity only captures lexical overlap and misses semantic contradictions (e.g., ``safe" versus ``unsafe"), we incorporate a natural language inference (NLI) model to refine similarity scores based on semantic relationships between sentences. This improved scoring better distinguishes knowledge-base-derived sentences from LLM-generated ones. Lastly, we conduct privacy-sensitive sentence classification by training a neural network that uses similarity features to detect sentences containing private knowledge base data.

This work represents the first attempt to comprehensively address the challenges of fine-grained privacy localization and multi-domain adaptability in RAG systems. Our study emphasizes the need for defenses that address both sentence-level attribution and cross-domain adaptability. In sum, our key contributions are as follows: 

\begin{itemize}
    \item We design a three-phase framework that decomposes adversarial queries, refines similarity scores using NLI, and classifies sentences to identify knowledge-base content. This allows us to accurately pinpoint the exact sentences containing private data, solving the ``which part leaks" problem that previous studies failed to address.
    \item We develop an adaptive query generation strategy that dynamically refines prompts to explore unknown domains. By combining broad initial queries with iterative refinement based on privacy-sensitive content, our method achieves robust performance on multi-domain datasets, which is previously unaddressed in RAG privacy research.
    \item  We conduct extensive experiments on diverse datasets and RAG configurations. The results demonstrate that our method outperforms baselines by over 30\% in key metrics such as ESR and F1-score, highlighting its effectiveness in real-world attack scenarios.
\end{itemize}

\section{Related Work}
Current research \cite{an2025rag,long2024backdoor, chen2024red,chaudhari2024phantom} shows that RAG systems have multiple security vulnerabilities, with related security studies primarily focusing on three attack vectors.

\textbf{Poisoning attacks} \cite{zou2024poisonedrag,ha2025mm,nazary2025poison,tan2024knowledge} inject malicious content into knowledge bases to corrupt LLM outputs. For example, Zou et al. \cite{zou2024poisonedrag} demonstrate this by making LLMs generate specific target answers, while Ha et al. \cite{ha2025mm} cause incorrect responses by introducing false information. In contrast to these attacks that require knowledge base access, our approach uses response comparison for fine-grained extraction, allowing it to work with closed RAG systems.


\textbf{Membership inference attacks} \cite{liu2024mask,shi2023detecting,anderson2024my,li2025generating} aim to determine if a document exists in the RAG knowledge base by analyzing response patterns. Liu et al. \cite{liu2024mask} infer document membership by masking parts of target documents and analyzing prediction accuracy. Shi et al. \cite{shi2023detecting} determine membership by calculating scores from least likely token sums. Anderson et al. \cite{anderson2024my} query RAG systems directly to detect if target documents appear in the retrieved context. Yet, they rely on having exact document copies, which is impractical for real-world private knowledge bases.


\textbf{Privacy extraction attacks} leverage adversarial prompts to make RAG systems reveal private data. For instance, Zeng et al. \cite{zeng2024good} and Qi et al. \cite{qi2024follow} design composite queries that force RAG systems to disclose private data from their knowledge bases. In a similar vein, Jiang et al. \cite{jiang2024rag} develop a proxy-based automated attack to extract large amounts of private information from these systems. Despite these efforts, existing privacy extraction attacks suffer from three critical limitations. First, they lack precision in identifying specific private content. Second, without knowledge base context, these attacks exhibit low efficiency and success rates. Finally, these methods have limited practical applicability due to insufficient cross-domain validation. Our work distinguishes itself by addressing these limitations through a knowledge-asymmetry-driven framework that enables precise extraction across multiple domains without requiring prior knowledge.

\section{Preliminary}
\subsection{Observable Knowledge Asymmetry in RAG Systems}
\label{S.Insight}

\textbf{Fundamentals of RAG.} The operation of a RAG system typically consists of three main stages. First, the system stores text data \emph{$T_D$ = \(\{T_1, T_2, \ldots, T_k\}\)}, which contains domain-specific knowledge and potentially private information, in the knowledge database \(\mathcal{D}\). This data is encoded into vector representations \(V_T\), forming the foundation for subsequent retrieval operations. When a user question \(\mathcal{Q}\) is received, the retriever \(\mathcal{P}\) encodes it into \emph{$V_Q$}. It then uses similarity measurements, such as cosine similarity, to compare \emph{$V_Q$} with stored \emph{$V_T$} vectors. This process retrieves the top-\emph{k} most relevant texts from \(\mathcal{D}\), expressed as: $Sim(V_{T},V_{Q})\Longrightarrow \{T_{1},T_{2},...,T_{k}\}\in \mathcal{D}$. Finally, an LLM \(\mathcal{M}\) integrates the retrieved contexts and the original question \(\mathcal{Q}\) to generate a more accurate answer \emph{$R_L$}, which can be expressed as: $R_{L}=\mathcal{M}(\{T_{1},T_{2},...,T_{k}\},\mathcal{Q})$.

\textbf{Response divergence.} RAG system responses depend on both LLM parameters \(\theta\) and retrieved knowledge \(\mathcal{T}_Q \subseteq \mathcal{D}\), whereas a standard LLMs (\(\mathcal{L}\)) generates responses \(A_L = \mathcal{L}(Q; \theta)\) using only \(\theta\). This produces a measurable content divergence \(\delta_Q = \Delta\left(\mathcal{M}(Q, T_Q; \theta), \mathcal{L}(Q; \theta)\right)\). The \(\Delta\) captures semantic or lexical differences, while \(\delta_Q\) quantifies how external knowledge creates unique asymmetries in RAG outputs, which our framework uses to identify sentences derived from the knowledge base. We experimentally validate this knowledge asymmetry phenomenon in Appendix \ref{appendixB}. These examples demonstrate how \(\delta_Q\) manifests in explicit content differences, with consistent patterns across medical, corporate , and multi-domain scenarios (Appendix \ref{appendixB}, Figures \ref{F.HCMExample}–\ref{F.NQExample}). RAG responses contain granular, context-specific details from external knowledge bases, whereas standard LLMs produce generic, pre-trained content. This insight inspires us to leverage \(\delta_Q\) as a diagnostic signal to isolate content uniquely originating from \(\mathcal{D}\), addressing the key challenge of separating private knowledge in private databases from LLM pre-trained information.

Our method identifies unique content from the external knowledge base \(\mathcal{D}\) by quantifying response divergence \(\delta_Q\), making it agnostic to specific private information types. This approach ensures that all knowledge-base content is detectable when it creates divergence, regardless of domain or category.

\subsection{Threat Model}

\textbf{Adversarial scenario.} We consider a fully black-box setting to mirror real-world API interactions where attackers only access the RAG system \(\mathcal{M}\) and a standard LLMs \(\mathcal{L}\) via public interfaces. Attackers lack access to internal components (knowledge database \(\mathcal{D}\), retriever \(\mathcal{P}\), LLM architecture \(\mathcal{A}\)) and metadata about \(\mathcal{D}\). They submit semantically meaningful adversarial queries \(Q_1, \dots, Q_m\), receiving text responses \(R_L = \mathcal{M}(Q)\) from the RAG system and \(A_L = \mathcal{L}(Q)\) from the standard LLMs without intermediate outputs, simulating real-world privacy inference via response comparison.

In this work, we consider two attack scenarios based on the attacker's knowledge of the application domain:

\textbf{(1)Single-Domain scenario.} The attacker possesses domain information related to the RAG knowledge base. For example, in vertical applications such as smart hospital consultant or financial report assistant, although the internal database is not visible, the domain information itself is public and known to any user.

\textbf{(2)Multi-Domain scenarios.} The attacker lacks prior knowledge of the domain information of the RAG knowledge base, such as in personal assistant applications. In this scenario, the RAG system's private knowledge base may contain documents spanning multiple domains. Therefore, how to design targeted adversarial queries without prior knowledge of domain information becomes a key challenge.

\begin{figure*}[t]
    \centering
    \setlength{\abovecaptionskip}{1.mm}
    \subfigure[RAG-Privacy on HCM]{
        \includegraphics[width=0.235\textwidth]{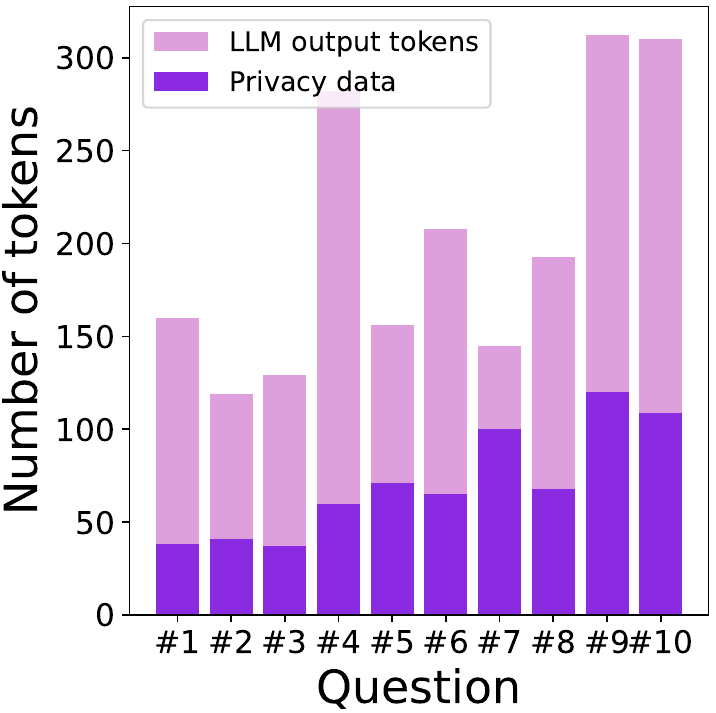}
    }
    \hspace{-0.01\textwidth}  
    \subfigure[RAG-Privacy on EE]{
        \includegraphics[width=0.235\textwidth]{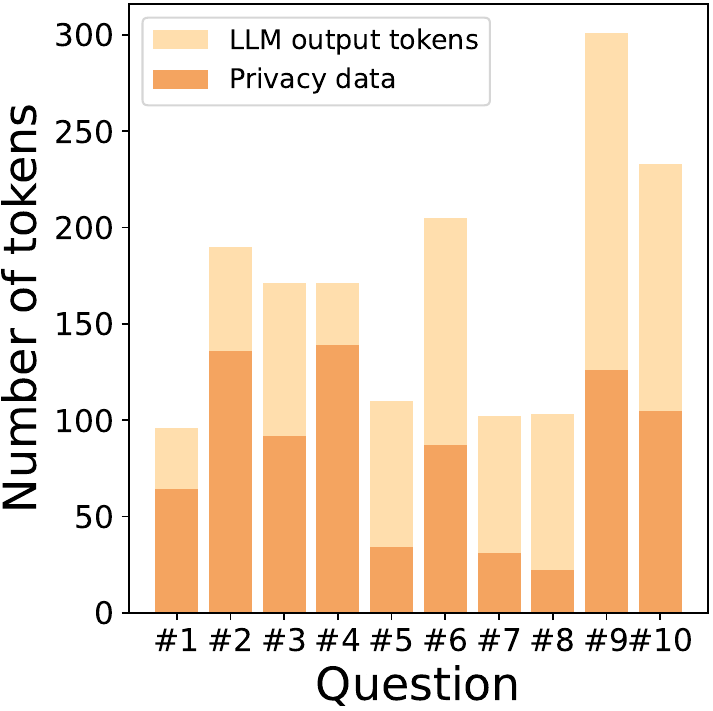}
    }
    \hspace{-0.01\textwidth}  
    \subfigure[RAG-Thief  on HCM]{
        \includegraphics[width=0.235\textwidth]{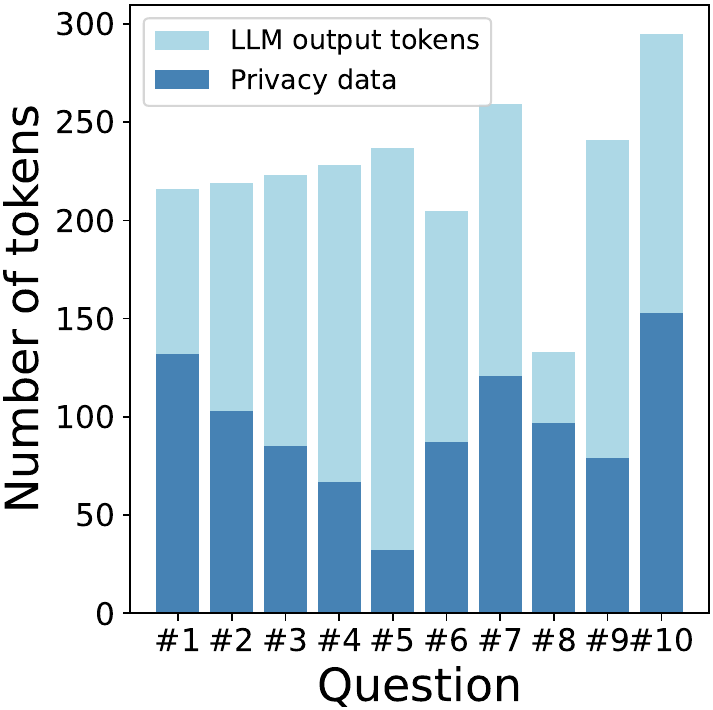}
    }
    \hspace{-0.01\textwidth}
    \subfigure[RAG-Thief on EE]{
        \includegraphics[width=0.235\textwidth]{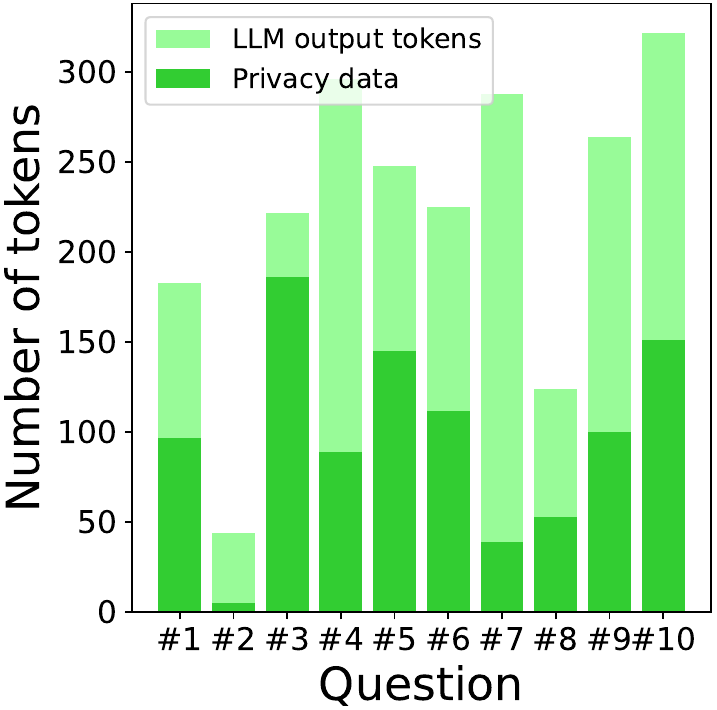}
    }
    \captionsetup{labelfont=bf}
    \caption{Proportion of private data in baseline attack outputs across datasets.}
    \label{Signalratio}
    \vspace{-5mm}
\end{figure*}

\textbf{Adversarial objectives.} Our attack targets two technical gaps in prior work:

\textbf{(1) Fine-grained privacy localization.} Existing privacy attacks on RAG systems, such as RAG-Privacy \citep{zeng2024good} and RAG-Thief \citep{jiang2024rag}, can lead to privacy leakage. However, they fail to distinguish between sentences derived from the private knowledge base \(\mathcal{D}\) and those originating from the LLM's pre-trained content in mixed responses. To verify this critical gap, we strictly reproduce their experimental setups while employing their specified RAG configurations. For both the HCM and EE datasets, we further develop 10 adversarial queries in accordance with their prompt templates. As shown in Figure \ref{Signalratio}, adversarial queries from prior work prompt RAG systems to generate responses containing private data, but the proportion of genuine private content is notably low, with non-private content dominating the output. More importantly, neither framework offers a effective method to separate these two content types. Owing to the information mixture problem where private knowledge \(\mathcal{T}_D\) blends with LLM pre-trained content, each sentence’s source becomes unidentifiable. This implies even if leakage occurs, users cannot pinpoint which specific sentence leaks private data, making these attacks unable to support targeted privacy mitigation. To address this issue, given a RAG system response $R_L = \{R_1, R_2, \dots, R_n\}$, we aim to precisely determine ${\arg\max}\mathds{1}(R_i \in \mathcal{T}_D)$ for each sentence \(R_i\) in \(R_L\), thus enabling fine-grained privacy data localization in RAG.

\textbf{(2) Cross-domain generalization.} Real-world RAG systems typically use multi-domain knowledge bases $\mathcal{D} = \bigcup_{d=1}^D \mathcal{D}_d$ that may combine $\mathcal{D}_1$ for medical records, $\mathcal{D}_2$ for financial reports, and $\mathcal{D}_3$ for legal documents. Current privacy attacks \cite{qi2024follow, zeng2024good,jiang2024rag} rely on single-domain knowledge, failing in heterogeneous environments. Our study aims to develop a unified framework for both single- ($\mathcal{D}=1$) and multi-domain ($\mathcal{D}\ge 2$) scenarios using domain-agnostic prompts and adaptive queries, overcoming reliance on preset patterns and enabling robust privacy evaluation across domains.

Knowledge overlap between \(\mathcal{D}\) and \(\mathcal{L}\) might reduce response divergence \(\delta_Q\) and affect content extraction, but such overlap constitutes public knowledge with no privacy leakage risk. Thus, we exclude it from consideration here, with further details discussed in Appendix \ref{appendixG}.


\section{Adversarial Attack Framework}

\begin{figure*}[t]
\centering
\includegraphics[width=1.0\textwidth]{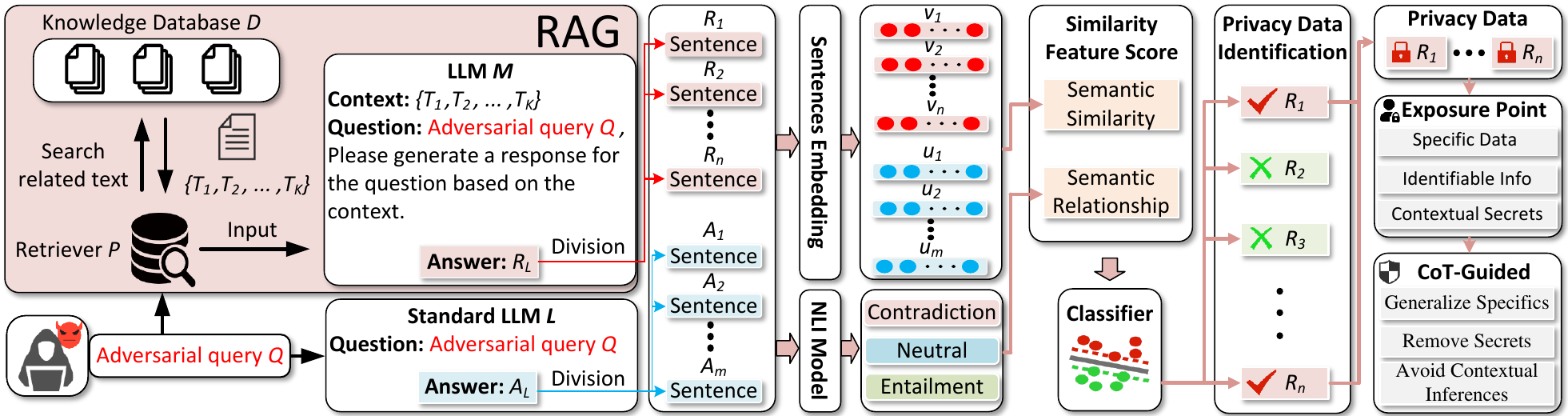}
\captionsetup{labelfont=bf}
\caption{Workflow of our attack on RAG system.}
\label{fig_1}                              
\end{figure*}

Figure \ref{fig_1} shows our three-phase attack framework: adversarial query generation, similarity feature score calculation, and privacy identification. Specifically, the attacker first submits adversarial query $\mathcal{Q}$ to both RAG system $\mathcal{M}$ and standard LLMs $\mathcal{L}$, obtaining responses \(R_L\) and \(A_L\). These responses are then segmented into sentence sets \(\{R_1, R_2, \ldots, R_n\}\) and \(\{A_1, A_2, \ldots, A_m\}\). Using sentence-embedding models, it generates fixed-dimensional vectors for these sets, while a natural language inference (NLI) model analyzes semantic relationships between sentences, classifying them as contradiction, neutral, or entailment. The system computes cosine similarity between the sentence-embedding vectors and adjusts these scores based on the NLI model's semantic classifications. Finally, a classifier uses these similarity feature scores to identify sentences containing private data within \(\{R_1, R_2, \ldots, R_n\}\). Afterward, we apply chain-of-thought reasoning to reconstruct the RAG outputs by generalizing and replacing sensitive details to generate privacy-preserving responses.


\subsection{Adversarial Query Generation}
\label{S.QueryGeneration}

Our attack framework's cornerstone lies in the generation of adversarial queries, a crucial step that sets it apart from existing methods. By splitting the query Q into two parts (\(\mathcal{Q} = q_1 \oplus q_2\), where \(\oplus\) represents text concatenation), we can manipulate the RAG system to reveal private data while maximizing the contrast between RAG and standard LLM responses, thereby facilitating the identification of knowledge base content.


\textbf{Maximizing informative responses with \(q_1\).} For single-domain knowledge bases, we employ a structured, open-ended question template for \(q_1\). This template is designed to incorporate keywords likely to appear in the knowledge base, such as \textit{"heart failure"}, \textit{"stroke"}, and \textit{"liver cirrhosis"} in a medical knowledge base. The query is formulated as :

\(\boldsymbol{q_1}\)\textbf{:} \textit{"Please tell me some information related to [keywords]."}

This open-ended format encourages both the RAG system and standard LLMs to generate comprehensive responses, ensuring that response differences reflect knowledge base access rather than variations in response length. When the RAG system encounters such queries (examples can be found in Appendix \ref{appendixE.1}), it retrieves and generates content based on the knowledge base, incorporating specific details such as patient histories or treatment protocols. In contrast, standard LLMs rely solely on their pre-trained corpus, producing generic and context-agnostic responses. As shown in Figure \ref{F.HCMExample} of Appendix \ref{appendixB}, their responses display systematic differences: the RAG system outputs patient-specific symptoms, while standard LLMs provide general medical knowledge applicable to the broader population. This response divergence stems directly from the knowledge asymmetry between the two models, which forms the foundation of our privacy extraction method.

In multi-domain settings with limited background knowledge, identifying effective keywords in the initial query \(q_1\) that trigger RAG systems to retrieve private data poses a challenge. We address this using an iterative refinement approach detailed in Algorithm~\ref{A.Query}.
First, we design a prompt template (Figure \ref{fig_5} of Appendix \ref{appendixE.2}) for an LLM to generate 10 broad, cross-domain initial questions (i.e., initial \(q_1\)). Examples of these initial questions are illustrated in Appendix \ref{appendixE.2}. We combine these questions with the pre-defined \(q_2\) to create the adversarial query set \(\mathcal{Q}\), which we input into both the RAG system and a standard LLM to collect responses. Using the similarity scoring described in Section \ref{S.similarity}, we compute feature scores for each sentence in the RAG responses. The classifier from Section \ref{S.Classifier} then extracts potential private data based on these scores. When initial \(q_1\) successfully triggers privacy leakage in RAG outputs, we optimize the initial queries by integrating the extracted privacy features (such as domain-specific keywords or semantic patterns). This creates a refined \(q_1\) that more precisely targets knowledge-base-specific content, leading the RAG system to retrieve and disclose more private data from the knowledge base. See Appendix \ref{appendixE.2} for related examples.

\begin{algorithm}[H]
\label{A.Query}
\small
\caption{Generation of \(q_1\) in multi-domain scenarios.}
\KwIn{An initial set of \(q_1\)=$\{q_1^{(1)}, q_1^{(2)}, \dots, q_1^{(10)}\}$ generated by an LLM; an attacker-chosen RAG system $\mathcal{M}$ and a standard LLM $\mathcal{L}$; a privacy classifier \emph{DNN}
}
\KwOut{A refined set of \(q_1\)}

\For{$i = 1$ \KwTo $10$}{
\(\mathcal{R}_i=\mathcal{M}(q_1^{(i)} \oplus  q_2)\); \(\mathcal{A}_i=\mathcal{L}(q_1^{(i)} \oplus q_2)\); $(\mathcal{S}_1, \mathcal{S}_2, \dots, \mathcal{S}_n) = \texttt{SimilarityScoring}(\text{Split}(\mathcal{R}_i), \text{Split}(\mathcal{A}_i))$

    \For{$j = 1$ \KwTo $n$}{
        \If{$\text{DNN}(\mathcal{S}_j) \Rightarrow \text{Contains private data}$}{
            $\hat{q}_1^{(i)} = \texttt{Add}(q_1^{(i)}, \text{private data})$ 
        }
    }
    \Return{\(\{\text{Refined}(q_1^{(i)})=\hat{q}_1^{(i)}\mid i = 1,\cdots,10\}\)}
}
\end{algorithm}

\textbf{Enhancing response divergence with \(q_2\).} To further amplify the differences between RAG and standard LLM responses, \(q_2\) is designed as an explicit prompt:

\(\boldsymbol{q_2}\)\textbf{:} \textit{"and provide contextual information based on the retrieved content."}


This design prompts the RAG system to retrieve relevant document fragments from its knowledge base and incorporate contextual knowledge when generating responses, enabling full use of proprietary knowledge during response generation. Unlike RAG, standard LLMs lack retrieval mechanisms and can only generate responses from their pre-trained corpus. By increasing the knowledge-base dependency in RAG outputs, \(q_2\) makes private data more likely to appear in \(R_L\) while \(A_L\) remains limited to generic pre-trained content. The expanded differences in semantics and content between the two response types allow for better identification and separation of knowledge-base sentences in RAG responses, leading to more accurate and efficient privacy extraction.

\subsection{Similarity Feature Score Calculation}
\label{S.similarity}

The core of our methodology centers on precisely extracting sentences containing private data from RAG system responses, which combine knowledge-base content and pre-trained LLM knowledge. To distinguish these two information sources, we introduce a similarity-based discrimination mechanism that leverages the knowledge asymmetry between RAG systems and standard LLMs. The goal of calculating semantic similarity is to identify sentences in \(R_L\) that deviate significantly from the LLM’s inherent knowledge (encoded in \(A_L\)), as such deviations are indicative of content originating from the external knowledge base \(\mathcal{D}\).

\textbf{Sentence-level semantic alignment analysis.} We first divide \(R_L\) and \(A_L\) into sentence sets \(\{R_1, R_2, \ldots, R_n\}\) and \(\{A_1, A_2, \ldots, A_m\}\) by using punctuation marks to separate sentences. This detailed segmentation allows for fine-grained comparison. Using the sentence embedding model, we then transform these sets into fixed-dimensional vectors \(\{v_i\}\) and \(\{u_j\}\) to enable numerical similarity calculations. For each sentence \(R_i\), we calculate the maximum cosine similarity \(S_i\) against all sentences \(A_j\), expressed as: 
\[
S_i = \max_{j \in [1, m]} \text{Cosine-sim}(v_i, u_j), \quad \forall i \in [1, n].
\]
This step quantifies the closest semantic match between \(R_i\) and any sentence in the LLM’s response. Low \(S_i\) values suggest that \(R_i\) contains information not present in the LLM's pre-trained corpus, indicating potential knowledge-base-derived private data. Conversely, high \(S_i\) values show alignment with general LLM knowledge, suggesting lower privacy leak risk.

\textbf{Mitigating limitations of cosine similarity.} Cosine similarity can capture surface-level semantic alignment but falls short with sentences that are structurally similar yet semantically opposite. For example, sentences like \textit{"this drug is safe"} and \textit{"this drug is unsafe"} would receive an artificially high similarity score because they share almost identical vocabulary, even though their meanings are opposite. To address this, we apply a natural language inference (NLI) model that classifies the semantic relationship between each \(R_i\) and its closest matching \(A_j\) (i.e., the one that maximizes \(\text{Cosine-sim}(v_i, u_j)\)) into three categories: contradiction, neutral, or entailment. The NLI model outputs a logits vector \(logits_{i,j} = [l_c, l_n, l_e]\), where each value represents the raw confidence score for its corresponding semantic relationship. The similarity score \(S_i\) is adjusted as follows:


$\bullet$ For contradictory classifications (\(\arg\max(logits_{i,j}) = l_c\)), we subtract \(l_c\) from \(S_i\) (\(\hat{S_i} = S_i - l_c\)) to penalize syntactic similarity between semantically conflicting sentences.


$\bullet$ For neutral relationships (no clear semantic connection), the score stays unchanged: \(\hat{S_i} = S_i\). Low cosine similarity appropriately indicates minimal knowledge overlap between responses.


$\bullet$ For entailment (when \(A_j\) logically implies \(R_i\)), we add \(l_e\) to increase similarity: \(\hat{S_i} = S_i + l_e\). This boosts scores for sentences aligning with LLM knowledge versus knowledge base content.


The resulting \(\hat{S_i}\) serves as the similarity feature score for sentences in \(\{R_1, R_2, \ldots, R_n\}\). This score combines both surface-level and deep semantic relationships, going beyond basic cosine similarity to precisely identify sentences that may pose privacy risks. Experiments in Appendix \ref{appendix2} demonstrate that \(\hat{S_i}\) effectively reflects the differences between private and non-private content.

\subsection{Privacy Sentence Identification}
\label{S.Classifier}
Following the computation of similarity feature scores, we frame privacy extraction as a binary classification task to identify sentences in \(R_L\) that contain private data from the knowledge base. We construct a dataset by pairing each sentence's similarity feature score with a manually annotated binary label. The annotation process works as follows: given a sentence set \(\{R_1, R_2, \ldots, R_n\}\) and retrieved top-\emph{k} text set \(\{T_1, T_2, \ldots, T_k\}\), we examine each generated sentence \(R_i\). If \(R_i\) appears in any retrieved text \(T_j\), we label its similarity feature score \(S_i\) with \(y_i = 1\); if it does not appear in any top-\emph{k} texts, we assign \(y_i = 0\). This can be expressed as:
\[
y_i = 
\begin{cases}
1, & \text{if } \exists j \in \{1, 2, \ldots, k\},\ R_i \in T_j \\
0, & \text{otherwise}
\end{cases}
\quad i \in \{1, 2, \ldots, n\}
\]
Using this dataset, we train a DNN classifier to map similarity features to privacy labels, enabling automated detection of privacy-sensitive sentences. This final classification stage allows our framework to precisely identify and extract knowledge-base content from RAG system responses.

\section{Evaluation}
\subsection{Experimental Setup}
\label{S.Setup}
\textbf{Dataset.} We use three representative datasets to simulate real-world privacy risks across different scenarios: HealthCareMagic (HCM) \citep{Health}, a single-domain medical corpus containing over 100,000 doctor-patient dialogues; Enron Email (EE) \citep{Enron}, a single-domain corporate dataset containing 500,000 employee emails; and NQ-train\_pairs (NQ) \citep{NQ}, a multi-domain benchmark dataset covering law, finance, and healthcare, containing over 30,000 question pairs to evaluate cross-domain generalization capabilities.

\textbf{RAG configuration.} We construct a modular RAG system using the LangChain framework, with components defined as follows: (1) \textit{Knowledge database.} We use HCM and EE datasets for single-domain attack validation, and NQ for multi-domain robustness testing. (2) \textit{Retriever.} Using three state-of-the-art dense retrievers (bge-large-en \citep{chen2024bge}, e5-large-v2 \citep{wang2022text}, and gte-large \citep{zhang2024mgte}), the RAG system calculates similarity through dot products between embeddings and employs FAISS with HNSW index \citep{douze2024faiss} to retrieve the top 3 most relevant texts as query context. (3) \textit{LLM backend.} We select \verb|LLaMA3.1-8B| \citep{dubey2024llama}, \verb|Qwen3-8B| \citep{yang2025qwen3}, and \verb|GPT-4o| \citep{achiam2023gpt} to evaluate cross-model generalization. These models cover a diverse range of popular commercial and open-source options.

\textbf{Attack framework setup.} The attack framework consists of three components. (1) \textit{Sentence embedding.} We use the \verb|all-MiniLM-L6-v2| model \citep{MiniLM} to convert sentences into 384-dimensional semantic vectors, allowing for numerical comparison of sentence meanings through cosine similarity. (2) \textit{Semantic relationship modeling.} We employ the \verb|Deberta-v3-large-mnli| model \citep{manakul2023selfcheckgpt} to analyze the semantic relationship between RAG and LLM responses. (3) \textit{Privacy classification.} A neural network \citep{hinton2006reducing}  trained on annotated data uses ReLU activation to classify similarity features for detecting private data.

\textbf{Data collection.} Following the adversarial query generation strategies detailed in Section \ref{S.QueryGeneration}, we design 30 adversarial queries for each knowledge database. We input these queries into both the RAG system and a standard LLMs to obtain responses. Using the method described in Section \ref{S.Classifier}, we created an annotated dataset, which we divided into training and test sets in a 7:3 ratio for model training and performance evaluation. Unless otherwise specified, our default setting uses \verb|LLaMA3.1-8B| for both the RAG system's generation model and the standard LLM, \verb|bge-large-en| as the retriever, and a temperature coefficient of 0.9. 

\textbf{Evaluation metrics.} To assess the performance of our approach, we use three key evaluation metrics. The extraction success rate (ESR) measures the proportion of correctly extracted private data from all actual private data, directly showing how well the model identifies and extracts sensitive information. The F1-score, combining precision and recall, provides a balanced assessment of the model's performance by showing its ability to detect private data while minimizing errors. Finally, the area under the curve (AUC) measures the model's classification ability, with higher values showing better distinction between private and non-private data across different thresholds.

\textbf{Baselines.} Existing works assess RAG privacy leaks by counting exposed data chunks \citep{zeng2024good,jiang2024rag}, whereas our approach identifies specific private content from the knowledge base in responses. This difference makes existing evaluation metrics unsuitable. Thus, we utilize the regex method mentioned in \citep{jiang2024rag} to extend existing works to our scenario. Additionally, we designed two tailored baselines based on \verb|GPT-4o| to evaluate our approach:


(1) \textit{Regex-based privacy identification} integrates the target private data types for extraction from RAG-Privacy \citep{zeng2024good} RAG-Thief \citep{jiang2024rag}, and designs regular expressions to extract specific types of private data from responses generated by RAG systems. Examples of these regular expressions are provided in Figure \ref{fig_16} in Appendix \ref{appendixI}.


(2) \textit{Content-based privacy discrimination} leverages LLM to detect whether RAG response sentences contain private information directly, where the prompt is shown in Appendix \ref{appendixH}. This baseline functions as a feature-driven privacy detector that assesses text based on explicit privacy indicators (e.g., age and diagnosis terms), instead of using knowledge asymmetry like our method.

(3) \textit{LLM-based privacy judgment} compares RAG system responses with standard LLM outputs and uses LLMs for analytical reasoning. Via carefully crafted prompts (Appendix \ref{appendixH}), it guides LLMs to analyze differences between these response sets, identifying content from external knowledge bases. This analysis of knowledge asymmetry aids in detecting private information within RAG outputs. Appendix \ref{appendixI} provides examples of privacy data extraction results across datasets using the baselines.

\subsection{Main Results}

\textbf{Overall performance.} Table \ref{table:llms_rag} shows our attack's performance across different datasets and LLMs. In our experiments, both the RAG system and standard LLM use the same generation model. For single-domain contexts (HCM and EE), our method consistently achieves ESRs above 90\% across various LLMs, demonstrating exceptional precision in extracting private data from domain-specific knowledge bases. In multi-domain scenarios (NQ) where other approaches falter, our method maintains an ESR of around 80\%, demonstrating the efficacy of our iterative prompt optimization strategy that dynamically targets emerging privacy features across diverse knowledge landscapes. 

\begin{wraptable}{r}{0.50\textwidth}
    \centering
    \setlength{\abovecaptionskip}{2mm}
    \small
    \vspace{-10pt}
    \caption{Overall performance.}
    \renewcommand{\arraystretch}{0.9} 
    \setlength{\tabcolsep}{3pt}
    \label{table:llms_rag}
    \begin{adjustbox}{width=\linewidth}
    \begin{tabular}{c | c c c c}
    \toprule
    \textbf{Datasets} & \textbf{LLMs of RAG} & \textbf{ESR} & \textbf{F1-Score} & \textbf{AUC} \\
    \midrule
    \multirow{3}{*}{HCM} & \verb|LLaMA3.1-8B| & 93.55\% & 92.06\% & 89.40\% \\
                          & \verb|Qwen3-8B|  & 90.91\% & 85.11\% & 84.30\% \\
                          & \verb|GPT-4o|    & 92.86\% & 96.30\% & 95.24\% \\
    \cmidrule(lr){1-5}
    \multirow{3}{*}{EE}   & \verb|LLaMA3.1-8B| & 95.65\% & 95.65\% & 91.30\% \\
                          & \verb|Qwen3-8B|  & 94.44\% & 87.18\% & 96.76\% \\
                          & \verb|GPT-4o|    & 90.91\% & 86.96\% & 95.45\% \\
    \cmidrule(lr){1-5}
    \multirow{3}{*}{NQ}    & \verb|LLaMA3.1-8B| & 80.00\% & 84.21\% & 86.67\% \\
                          & \verb|Qwen3-8B|  & 87.50\% & 84.85\% & 90.81\% \\
                          & \verb|GPT-4o|    & 76.73\% & 80.00\% & 84.59\% \\
    \bottomrule
    \end{tabular}
    \end{adjustbox}
    \vspace{-12pt}
\end{wraptable}

The F1-score remains above 80\% across all models, indicating balanced precision and recall in privacy discrimination. Additionally, our method achieves AUC values exceeding 84\% across all models, showing our framework's consistency and ability to generalize across heterogeneous datasets.
Our method performs better in single-domain scenarios than in cross-domain settings. This difference stems from the concentrated nature of single-domain data, where private information is tightly clustered around specific themes (e.g., cardiology in HCM, financial operations in EE). This focus allows our adversarial queries to trigger targeted knowledge-base retrievals, creating clear response disparities between RAG and standard LLMs.

\begin{wraptable}{r}{0.50\textwidth}
    \centering
    \setlength{\abovecaptionskip}{2mm}
    \vspace{-12pt}
    \caption{Comparison with baseline methods.}
    \renewcommand{\arraystretch}{0.6}
    \setlength{\tabcolsep}{3pt}
    \small 
    \label{table:baseline}
    \begin{tabular}{c | c c c}
    \toprule
    \textbf{Datasets} & \textbf{Methods} & \textbf{ESR} & \textbf{F1-Score} \\
    \midrule
    \multirow{5}{*}{HCM} & RAG-Privacy & 57.58\% & 46.15\%  \\
                         & RAG-Thief  & 61.37\% & 41.76\%  \\
                         & Content-based & 58.82\% & 51.28\% \\
                         & LLM-based & 65.22\% & 62.50\% \\
                         & \cellcolor{gray!20} \textbf{Ours} & \cellcolor{gray!20} \textbf{93.55\%} & \cellcolor{gray!20} \textbf{92.06\%} \\
    \cmidrule(lr){1-4}
    \multirow{5}{*}{EE}  & RAG-Privacy & 61.25\% & 43.64\%  \\
                         & RAG-Thief  & 52.75\% & 38.51\%  \\
                         & Content-based & 36.00\% & 47.37\% \\
                         & LLM-based & 60.87\% & 53.85\% \\
                         & \cellcolor{gray!20} \textbf{Ours} & \cellcolor{gray!20} \textbf{95.65\%} & \cellcolor{gray!20} \textbf{95.65\%} \\
    \cmidrule(lr){1-4}
    \multirow{5}{*}{NQ}  & RAG-Privacy & 37.25\% & 41.33\%  \\
                         & RAG-Thief  & 41.73\% & 33.64\%  \\
                         & Content-based & 18.75\% & 30.00\% \\
                         & LLM-based & 60.00\% & 43.90\% \\
                         & \cellcolor{gray!20} \textbf{Ours} & \cellcolor{gray!20} \textbf{80.00\%} & \cellcolor{gray!20} \textbf{84.21\%} \\
    \bottomrule
    \end{tabular}
    \vspace{-12pt}
\end{wraptable}

\textbf{Comparison with baselines.} Table \ref{table:baseline} quantifies the performance of our method against 4 baselines. The regex-based approach performs moderately in single-domain settings, with about 60\% ESR and low F1 scores (e.g., only 38.51\% on EE). This is due to its heavy reliance on specific data formats, leading to misidentification of non-private data and poor multi-domain adaptability. It only extracts via typical formats (e.g., names, phone numbers), resulting in sharply reduced performance (e.g., RAG-Privacy’s 37.25\% ESR on NQ).

Content-based discrimination also works moderately in single domains (e.g., 58.82\% ESR on HCM) by identifying explicit keywords, but its accuracy plummets in cross-domain settings (e.g., 18.75\% ESR on NQ) due to dependence on domain-specific features. The LLM-based judgment method improves on this via knowledge asymmetry, achieving better ESR on HCM (65.22\%) and NQ (60.00\%), but suffers from low precision (e.g., 43.90\% F1 on NQ) due to overgeneralization (misclassifying public information as private).

In contrast, our method consistently outperforms baselines across all datasets, achieving 93.55\% ESR on HCM, 95.65
\% on EE, and 80.00\% on NQ. The F1-Score exceeds baselines by 29-60\%, demonstrating high precision in extracting domain-specific private data. Moreover, via adaptive query refinement for unknown domains, it resolves low privacy extraction precision from insufficient context, while addressing baselines’ limitations in fine-grained detection and cross-domain generalization.

\vspace{-2mm}
\section{Ablation Study}
\vspace{-2mm}

\begin{table*}[h]
\centering
\begin{minipage}[t]{0.49\textwidth}
    \centering
    \setlength{\abovecaptionskip}{2mm}
    \small
    \caption{Impact of standard LLMs.}
    \renewcommand{\arraystretch}{0.97} 
    \setlength{\tabcolsep}{3pt}      
    \label{table:standard_llms}
    \begin{adjustbox}{width=\linewidth}
    \begin{tabular}{c | c c c c}
    \toprule
    \textbf{Datasets} & \textbf{Standard LLMs} & \textbf{ESR} & \textbf{F1-Score} & \textbf{AUC} \\
    \midrule
    \multirow{3}{*}{HCM} & \verb|LLaMA3.1-8B| & 93.55\% & 92.06\% & 89.40\% \\
                         & \verb|Qwen3-8B|     & 85.71\% & 82.76\% & 92.86\% \\
                         & \verb|GPT-4o|    & 88.89\% & 86.49\% & 85.12\% \\
    \cmidrule(lr){1-5}
    \multirow{3}{*}{EE}  & \verb|LLaMA3.1-8B| & 95.65\% & 95.65\% & 91.30\% \\
                         & \verb|Qwen3-8B|     & 88.24\% & 85.71\% & 94.12\% \\
                         & \verb|GPT-4o|    & 88.89\% & 81.00\% & 77.78\% \\
    \cmidrule(lr){1-5}
    \multirow{3}{*}{NQ}  & \verb|LLaMA3.1-8B| & 80.00\% & 84.21\% & 86.67\% \\
                         & \verb|Qwen3-8B|     & 77.78\% & 82.35\% & 87.04\% \\
                         & \verb|GPT-4o|    & 85.71\% & 80.00\% & 90.36\% \\
    \bottomrule
    \end{tabular}
    \end{adjustbox}
\end{minipage}
\hfill
\begin{minipage}[t]{0.49\textwidth}
    \centering
    \setlength{\abovecaptionskip}{2mm}
    \small
    \caption{Impact of different retrievers.}
    \renewcommand{\arraystretch}{1} 
    \setlength{\tabcolsep}{3pt}      
    \label{table:retriever_rag}
    \begin{adjustbox}{width=\linewidth}
    \begin{tabular}{c | c c c c}
    \toprule
    \textbf{Datasets} & \textbf{Retriever of RAG} & \textbf{ESR} & \textbf{F1-Score} & \textbf{AUC} \\
    \midrule
    \multirow{3}{*}{HCM} & \verb|bge-large-en| & 93.55\% & 92.06\% & 89.40\% \\
                         & \verb|e5-large-v2|  & 94.12\% & 91.43\% & 93.03\% \\
                         & \verb|gte-large|    & 94.44\% & 90.67\% & 95.83\% \\
    \cmidrule(lr){1-5}
    \multirow{3}{*}{EE}  & \verb|bge-large-en| & 95.65\% & 95.65\% & 91.30\% \\
                         & \verb|e5-large-v2|  & 82.14\% & 86.79\% & 93.28\% \\
                         & \verb|gte-large|    & 85.19\% & 85.19\% & 88.05\% \\
    \cmidrule(lr){1-5}
    \multirow{3}{*}{NQ}  & \verb|bge-large-en| & 80.00\% & 84.21\% & 86.67\% \\
                         & \verb|e5-large-v2|  & 84.62\% & 91.67\% & 95.97\% \\
                         & \verb|gte-large|    & 84.62\% & 88.00\% & 92.52\% \\
    \bottomrule
    \end{tabular}
    \end{adjustbox}
\end{minipage}
\end{table*}

\vspace{-3mm}

\textbf{Impact of standard LLMs.} Since our method relies on comparing responses generated by standard LLMs, differences in pre-training data between different LLMs may affect the results. In our experiments, we used \verb|LLaMA3.1-8B| as the generation model for RAG. Table \ref{table:standard_llms} demonstrates consistent performance across datasets with various standard LLMs, indicating that its effectiveness is not limited to any single language model. 

\textbf{Impact of retriever.} Different retrievers may influence how adversarial queries retrieve knowledge base content, affecting the level of privacy data exposure. As evidenced by consistent performance across three datasets in Table \ref{table:retriever_rag}, our method maintains effectiveness across different retrievers. 

\textbf{Impact of \(q_1\) and \(q_1\)\( \oplus \)\(q_2\).} 
By default, we construct adversarial queries by combining \(q_1\) and \(q_2\) into \(q_1\)\( \oplus \)\(q_2\). We studied the effectiveness when using only \(q_1\). The results in Table \ref{table:adv_queries} of Appendix \ref{appendixA} show that using \(q_1\) alone significantly reduces attack success rates, with ESR on HCM falling to only 55\%. Adding \(q_2\) increases ESR to 93.55\%, confirming the effectiveness of our \(q_2\) design.

\textbf{Impact of model size.} The size of a model's pre-training data affects the richness of its output content, which influences the response differences between RAG systems and standard LLMs. Experimental results in Figure \ref{modelsize} in Appendix \ref{appendixA} show that our attack method remains effective across models of different sizes under default settings, with most test models achieving ESR above 80\%.

\textbf{Impact of temperature.} Higher temperature increases response randomness, affecting knowledge base content in RAG outputs and experimental results. Figure \ref{fig_2} in Appendix \ref{appendixA} shows all metrics remain above 80\%, confirming robustness across temperature settings.

Appendix \ref{appendixD} discusses a defense strategy against the proposed attack. This strategy employs chain-of-thought reasoning with adaptive prompts to steer RAG systems away from sensitive content while experimentally proving its effectiveness. Appendix \ref{appendixF} covers the time cost and ethical considerations of our method. Appendix \ref{appendixG} explores the impacts of knowledge overlap and privacy techniques (e.g., differential privacy) on our approach.

\vspace{-2mm}
\section{Conclusion}
\vspace{-2mm}

We present a black-box attack framework for RAG systems that enables precise privacy localization and cross-domain generalization by exploiting knowledge asymmetry. Our method achieves up to 90\% ESR in single domains and 80\% ESR in multi-domain settings, surpassing baselines by over 30\% in key metrics. This work uncovers critical vulnerabilities in RAG systems, providing the first systematic solution for precise privacy extraction and establishing a benchmark for robust defenses in knowledge-augmented models.

\bibliographystyle{unsrt}  
\small
\bibliography{References}


\appendix

\newpage
\startcontents[appendix]
\printcontents[appendix]{}{1}{\section*{\centering Table of Contents for Appendix}}

\newpage
\section{Supplementary Ablation Experiment Results}
\label{appendixA}


\begin{table*}[htbp]
\centering
\scriptsize
\caption{Impact of Adversarial Queries.}
\label{table:adv_queries}
\setlength{\tabcolsep}{3pt}      
\renewcommand{\arraystretch}{1} 
\begin{adjustbox}{width=0.7\textwidth}
\begin{tabular}{c | c c c c}
\toprule
\textbf{Datasets} & \textbf{Adversarial Queries} & \textbf{ESR} & \textbf{F1-Score} & \textbf{AUC} \\
\midrule
\multirow{2}{*}{HCM} & \( q_1 \)            & 55.56\% & 66.67\% & 89.35\% \\
                     & \( q_1 \oplus q_2 \) & 93.55\% & 92.06\% & 89.40\% \\
\cmidrule(lr){1-5}
\multirow{2}{*}{EE}  & \( q_1 \)            & 81.30\% & 79.25\% & 60.87\% \\
                     & \( q_1 \oplus q_2 \) & 95.65\% & 95.65\% & 91.30\% \\
\cmidrule(lr){1-5}
\multirow{2}{*}{NQ}  & \( q_1 \)            & 52.94\% & 64.29\% & 77.73\% \\
                     & \( q_1 \oplus q_2 \) & 80.00\% & 84.21\% & 86.67\% \\
\bottomrule
\end{tabular}
\end{adjustbox}
\end{table*}

\begin{figure}[H]
    \centering 
    \subfigure[Extraction success rate]{
        \includegraphics[width=0.31\linewidth]{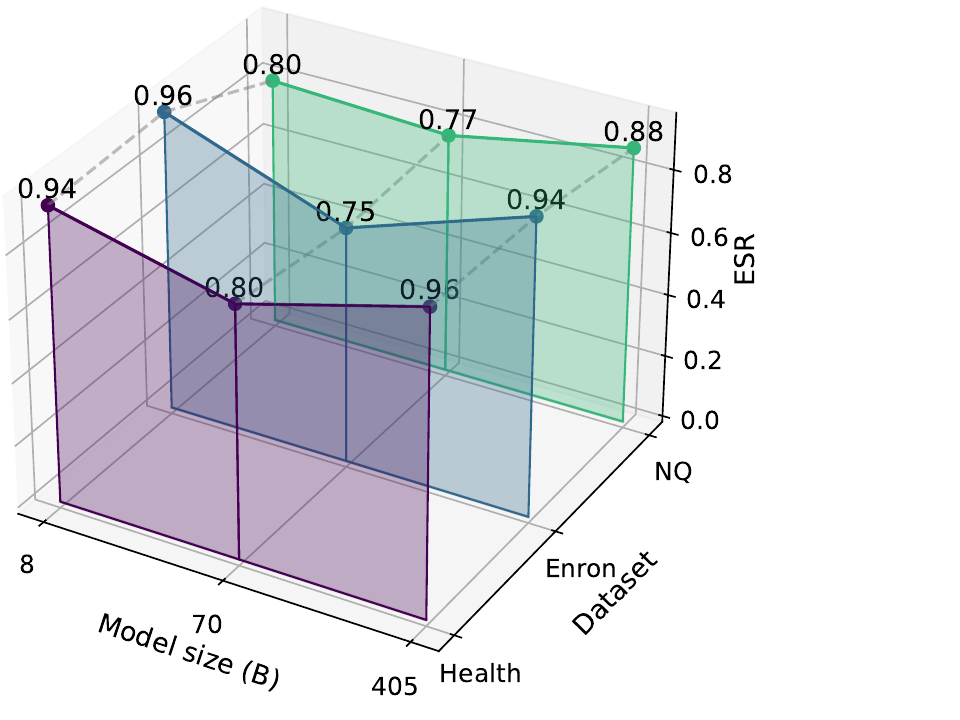}
    }
    \subfigure[F1-Score]{
        \includegraphics[width=0.31\linewidth]{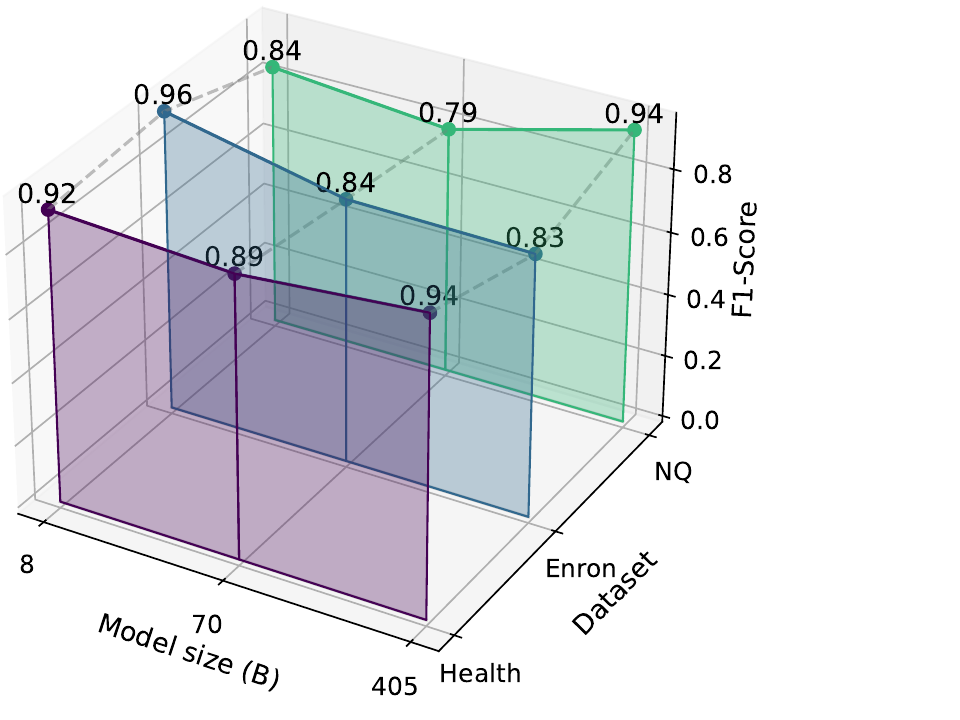}
    }
    \subfigure[AUC]{
        \includegraphics[width=0.31\linewidth]{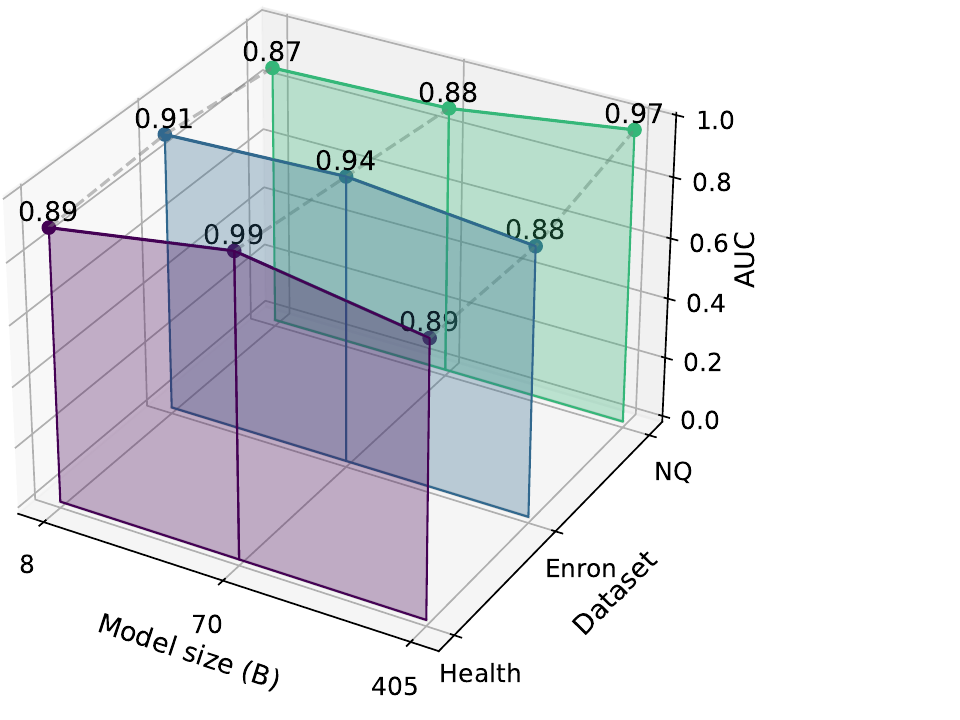}
    }
    \caption{Performance under different model size.}
    \label{modelsize}
\end{figure}
\vspace{-6mm}
\begin{figure}[h]
    \centering 
    \subfigure[Extraction success rate]{
        \includegraphics[width=0.31\linewidth]{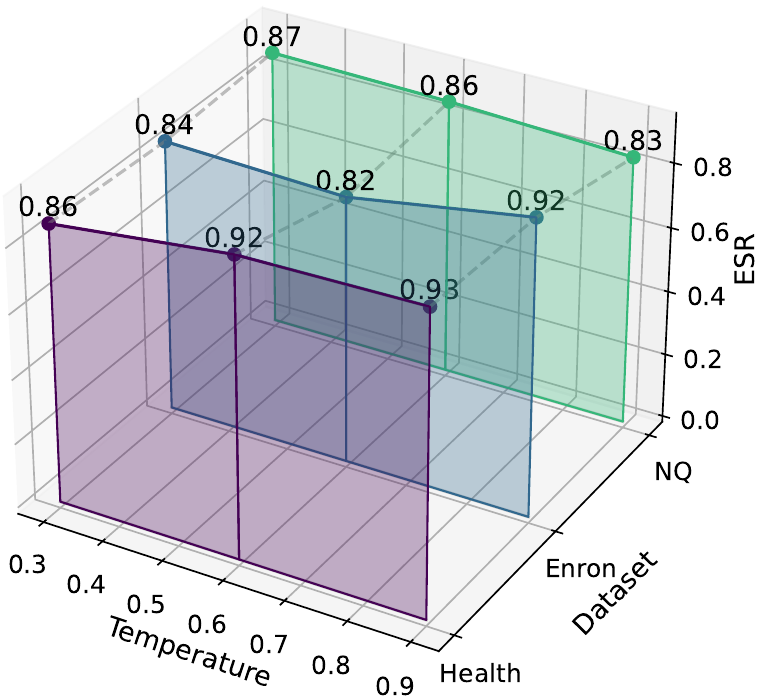}
    }
    \subfigure[F1-Score]{
        \includegraphics[width=0.31\linewidth]{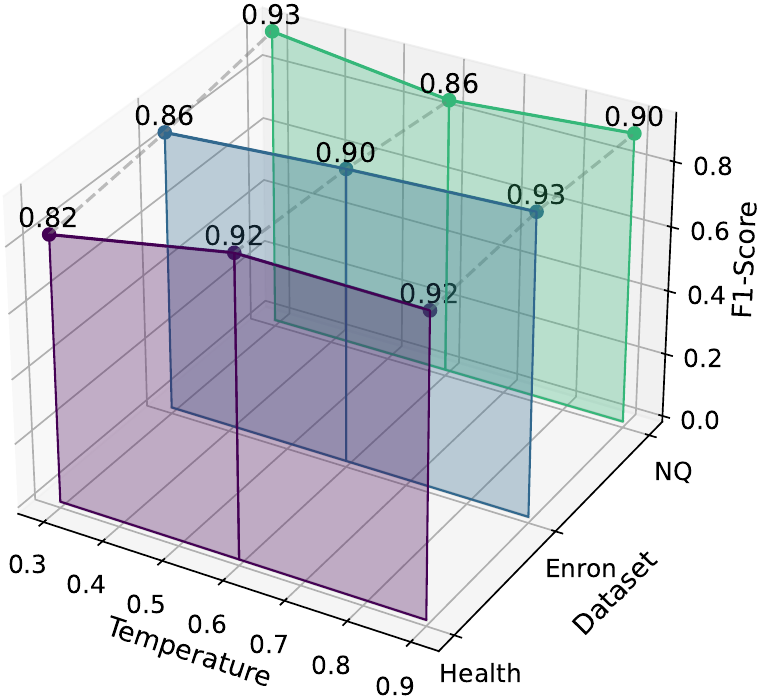}
    }
    \subfigure[AUC]{
        \includegraphics[width=0.31\linewidth]{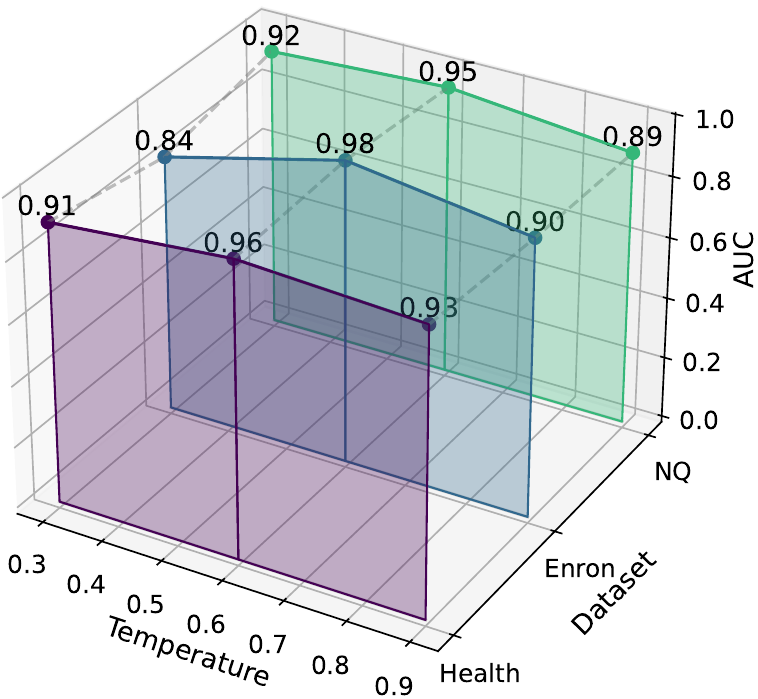}
    }
    \caption{Performance under different temperature.}
    \label{fig_2}
\end{figure}

\newpage
\section{Examples of Knowledge Asymmetry }
\label{appendixB}

\begin{figure}[H]
\centering
\includegraphics[width=1.0\textwidth]{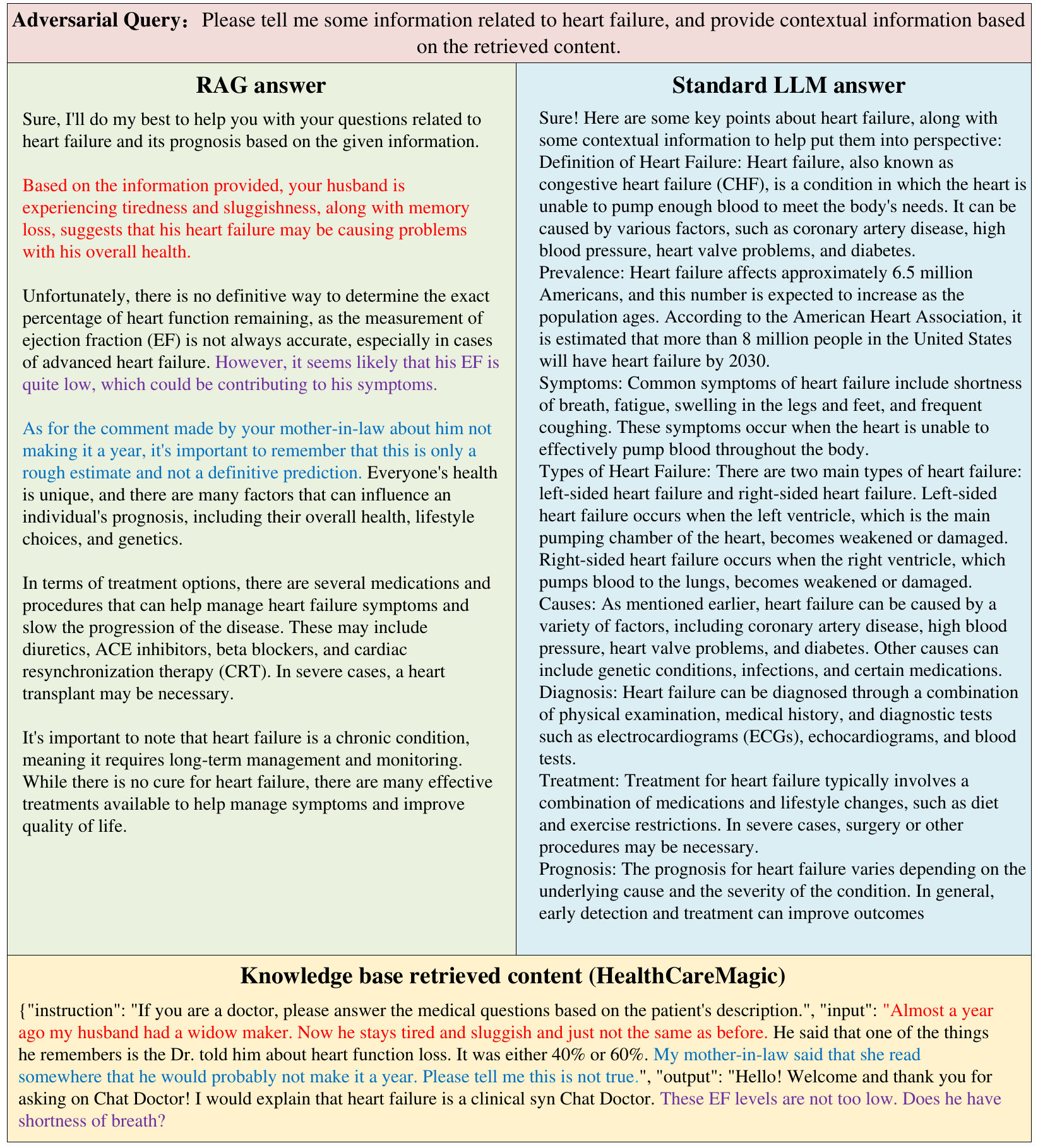}
\caption{Differences in responses on the HealthCareMagic database.}
\label{F.HCMExample}    
\end{figure}

Figure \ref{F.HCMExample} demonstrates that in the HealthCareMagic database, due to the knowledge asymmetry between RAG systems and standard large language models (LLMs), they generate different responses when faced with the same question \textit{"Please tell me some information related to heart failure, and provide contextual information based on the retrieved content"}. In the RAG system's answer, the color-coded sentences represent the privacy information we identified, which directly originates from the retrieved knowledge base text, with corresponding original data marked in matching colors.

For example, the RAG response \textit{"Based on the information provided, your husband is experiencing tiredness and sluggishness, along with memory loss, suggests that his heart failure may be causing problems with his overall health."} can be traced back to the knowledge base statement \textit{"Almost a year ago my husband had a widow maker. Now he stays tired and sluggish and just not the same as before"}. From this, we can infer that this woman's husband has heart disease, accompanied by symptoms such as persistent fatigue, lethargy, and memory decline. These contents represent the potential privacy information we extracted, demonstrating how context generation driven by retrieved content in specific RAG mechanisms may risk exposing users' sensitive data.

\begin{figure}[H]
\centering
\includegraphics[width=1.0\textwidth]{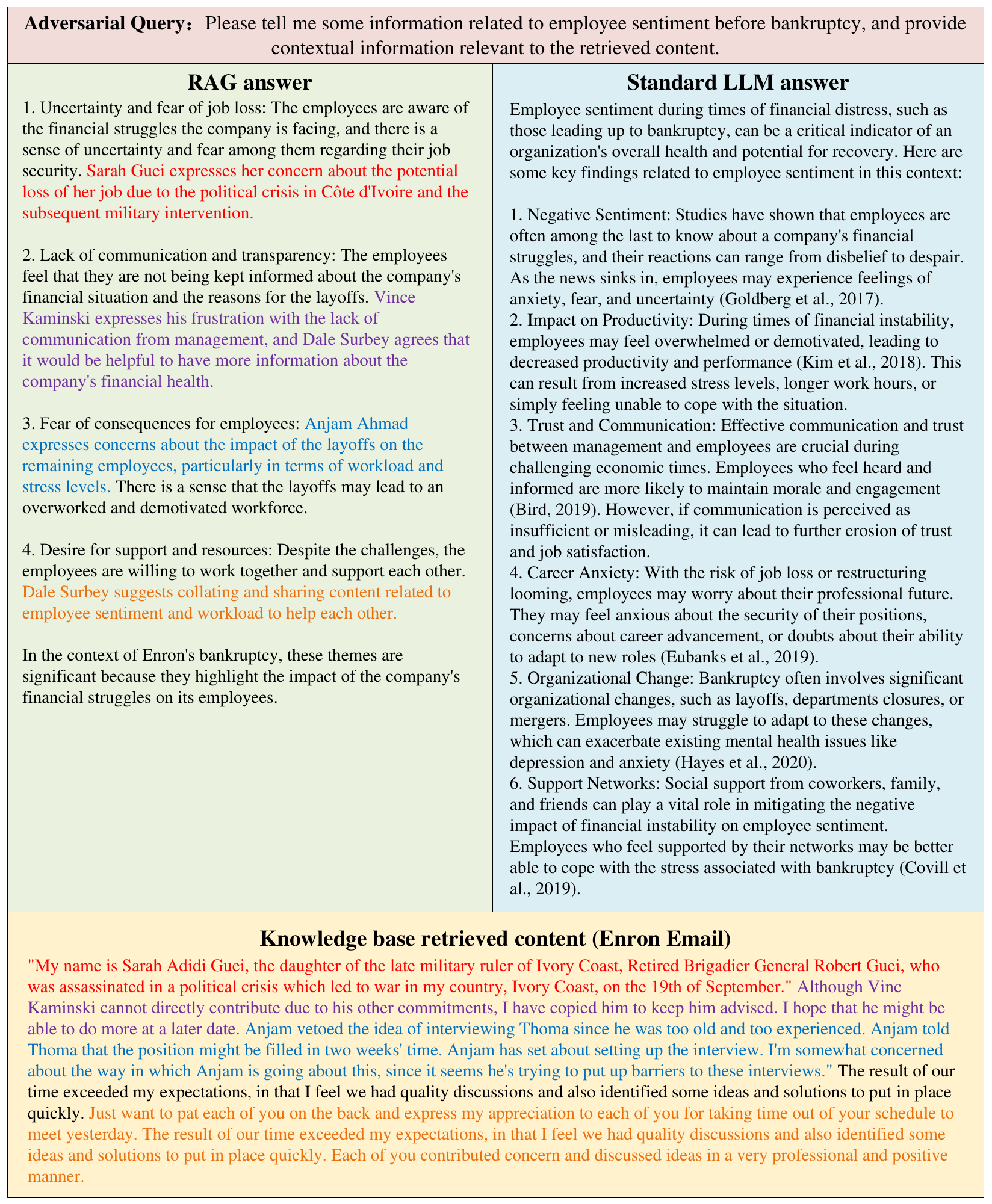}
\caption{Differences in responses on the Enron Email database.}
\label{F.EEExample} 
\end{figure}

Figure \ref{F.EEExample} presents a similar phenomenon in the Enron Email dataset, where the query \textit{"Please tell me some information related to employee sentiment before bankruptcy, and provide contextual information relevant to the retrieved content"} prompts RAG systems to retrieve and incorporate specific internal communications. 

One example is the sentence \textit{"Sarah Guei expresses her concern about the potential loss of her job due to the political crisis in Côte d'Ivoire..."}, which originates from a personal email containing the line \textit{"My name is Sarah Adidi Guei, the daughter of the late military ruler of Ivory Coast..."}. This case demonstrates that RAG systems not only retrieve private documents but may also expose sensitive individual identities, family backgrounds, and geopolitical associations embedded within them. Unlike standard LLMs, which tend to provide generalized summaries or aggregate sentiment analysis, RAG systems may reproduce verbatim fragments of confidential records, posing a concrete risk of privacy breaches in enterprise and corporate settings.

\begin{figure}[H]
\centering
\includegraphics[width=1.0\textwidth]{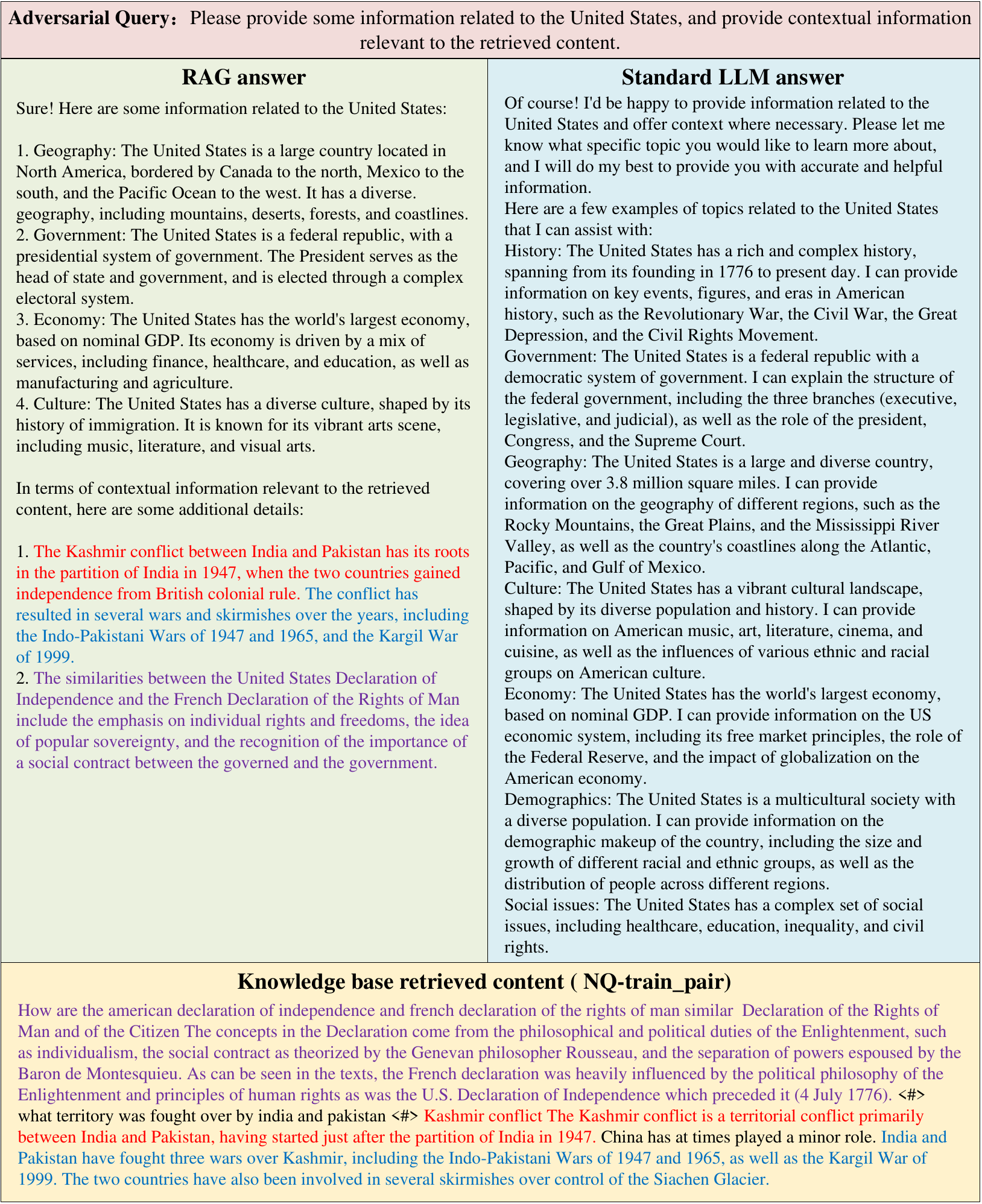}
\caption{Differences in responses on the NQ-train\_pair database.}
\label{F.NQExample} 
\end{figure}

Figure \ref{F.NQExample} shows how RAG systems and standard large language models (LLMs) respond differently to the question \textit{"Please provide some information related to the United States, and provide contextual information relevant to the retrieved content"} in the Natural Questions dataset. The RAG model provides fundamental information about the United States. In its "contextual information" section, the model introduces two additional topics: the Kashmir dispute between India and Pakistan, and parallels between the U.S. Declaration of Independence. All of these details directly match content found in the original knowledge base text.

While the above content does not directly contain personal information or explicit privacy data, it highlights another significant issue: when handling localized or topic-specific queries, RAG systems may pull irrelevant content from multi-domain databases, causing semantic drift. This drift not only makes responses less specific and accurate but can also lead to cross-contextual content leakage. In multi-domain database environments, such unexpected information exposure could reveal patterns in the original knowledge sources' structure and content distribution, creating risks of indirect privacy leaks.

\newpage
\section{Examples of Privacy Data Precise Extraction}
\label{appendix2}

\begin{figure}[h]
\centering
\includegraphics[width=1.0\textwidth]{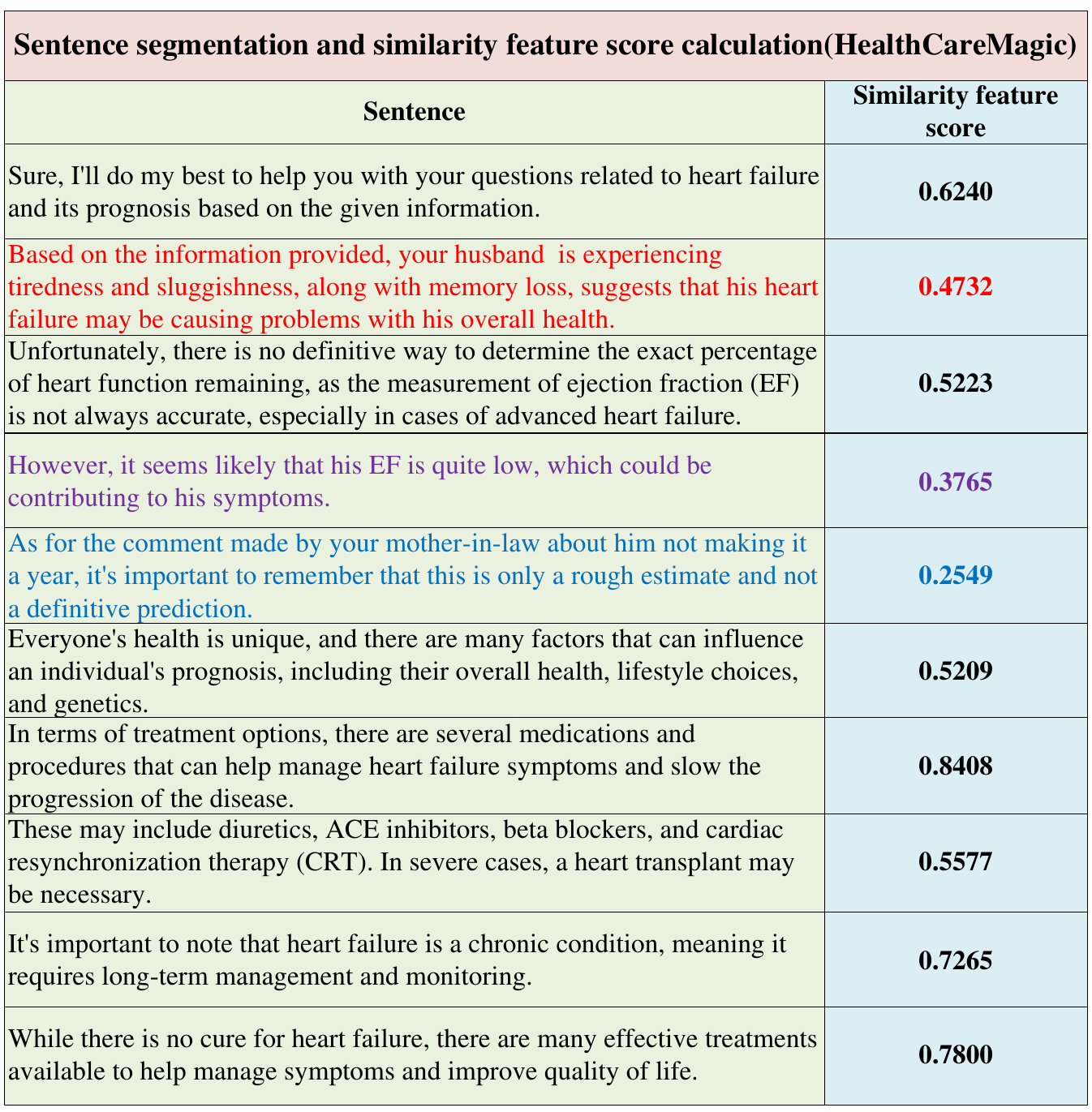}
\caption{Sentence segmentation and similarity scores \(\hat{S_i}\) on HealthCareMagic.}
\label{fig_4}                               
\end{figure}

\begin{figure}[h]
\centering
\includegraphics[width=1.0\textwidth]{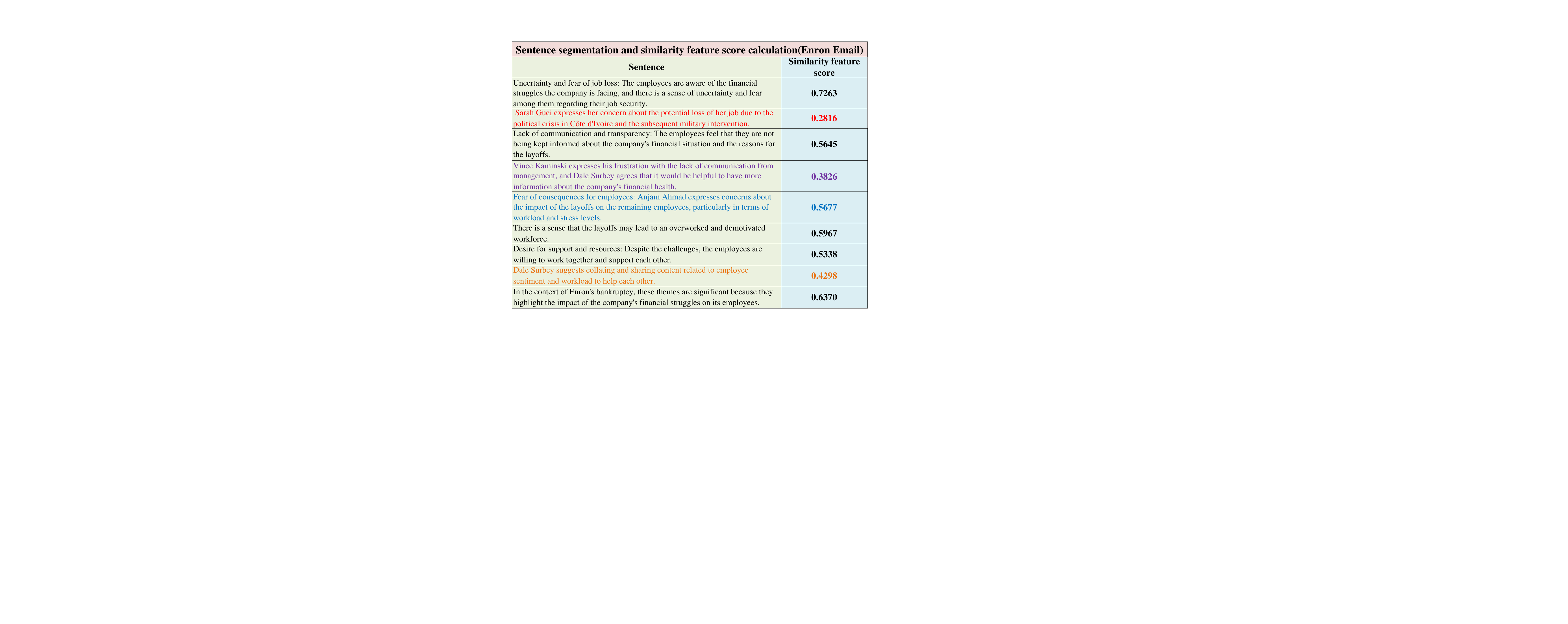}
\caption{Sentence segmentation and similarity scores \(\hat{S_i}\) on Enron Email.}
\label{Similarity Feature Score2}                     
\end{figure}

\begin{figure}[H]
\centering
\includegraphics[width=1.0\textwidth]{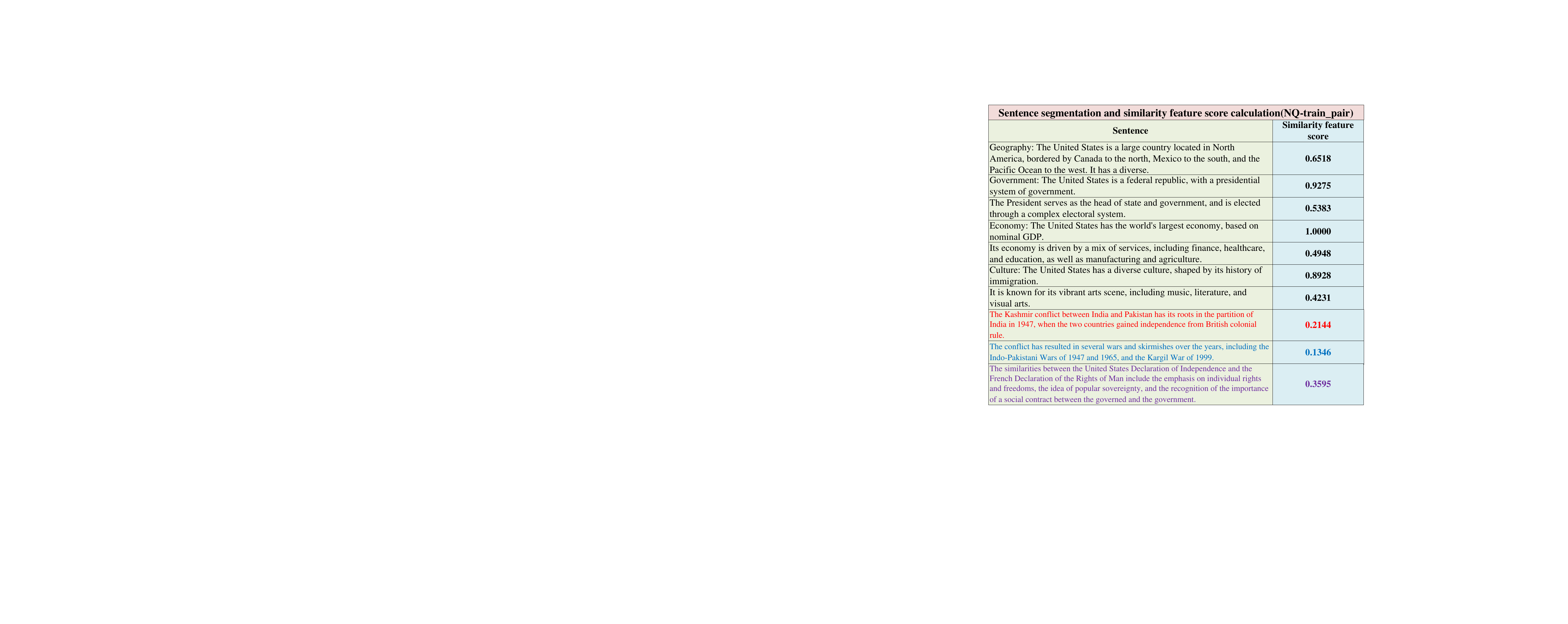}
\caption{Sentence segmentation and similarity scores \(\hat{S_i}\) on NQ-train\_pair.}
\label{Similarity Feature Score3}               
\end{figure}

Figure \ref{fig_4}, Figure \ref{Similarity Feature Score2}, and Figure \ref{Similarity Feature Score3} demonstrate examples of our experimental results on sentence segmentation and similarity feature score calculation for RAG-generated responses across multiple datasets. The color-highlighted sentences represent identified private content that directly originates from the retrieved knowledge base and cannot be generated by standard large language models (LLMs) without access to such external data—highlighting the knowledge asymmetry between RAG systems and standard LLMs.

These examples show that due to the unique source and contextual differences of private content, sentences containing private information tend to have notably lower similarity feature scores compared to sentences without privacy content, demonstrating clear separability. For instance, in the HealthCareMagic example shown in Figure \ref{fig_4}, the similarity feature scores of privacy-related sentences are 0.4732, 0.3765, and 0.2549, while the lowest similarity score among non-privacy sentences is 0.5223—higher than the highest score among the privacy sentences. This difference proves that similarity feature scores can effectively identify potential private data in RAG responses.

\newpage
\section{Defense method}
\label{appendixD}
\subsection{Privacy-preserving response generation}
\label{appendixD.1}
Building on precise detection of privacy-sensitive sentences, we introduce a privacy-preserving response generation strategy that leverages chain-of-thought (CoT) reasoning to refactor RAG outputs, preventing RAG systems from revealing sensitive content. This operates in two stages, designed to universally handle sensitive information by focusing on generic privacy risks.

\textbf{Exposure point analysis.} For each identified privacy-sensitive sentence \(R_i \in R_L\) with low \(\hat{S_i}\), we employ a domain-agnostic taxonomy to analyze their semantic context and detect whether they contain: (1) Exact values or details that uniquely identify entities (e.g., account balances, IDs). (2) Personal or corporate confidential information (e.g., PII, internal memos). (3) Sensitive content derived from context (e.g., user preferences, location data). 

\textbf{CoT-guided response refactoring.} Based on exposure point analysis, we design a CoT prompt to guide the RAG system in reformulating responses while balancing utility and privacy. First, we generalize specifics by replacing sensitive data, including exact values and identifiers, with semantically equivalent generalizations that maintain the original meaning while protecting privacy. Second, we implement structured reasoning chains that guide the RAG system to explain conclusions using domain-agnostic frameworks instead of raw private data. 
\subsection{Case Study for Privacy-Preserving Response Generation}
\label{cot}

\begin{figure}[h]
\centering
\includegraphics[width=1.0\textwidth]{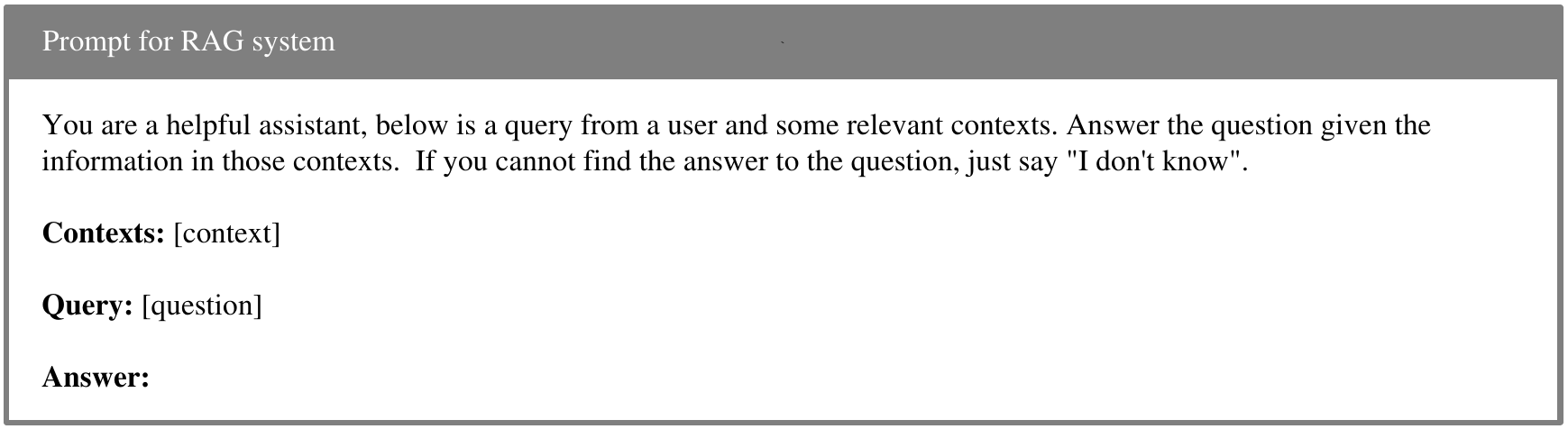}
\caption{Prompt for RAG system.}
\label{COT(1)}               
\end{figure}

Figure \ref{COT(1)} shows the system prompt used in RAG to let a LLM generate an answer based on the given context.

\begin{figure}[H]
\centering
\includegraphics[width=1.0\textwidth]{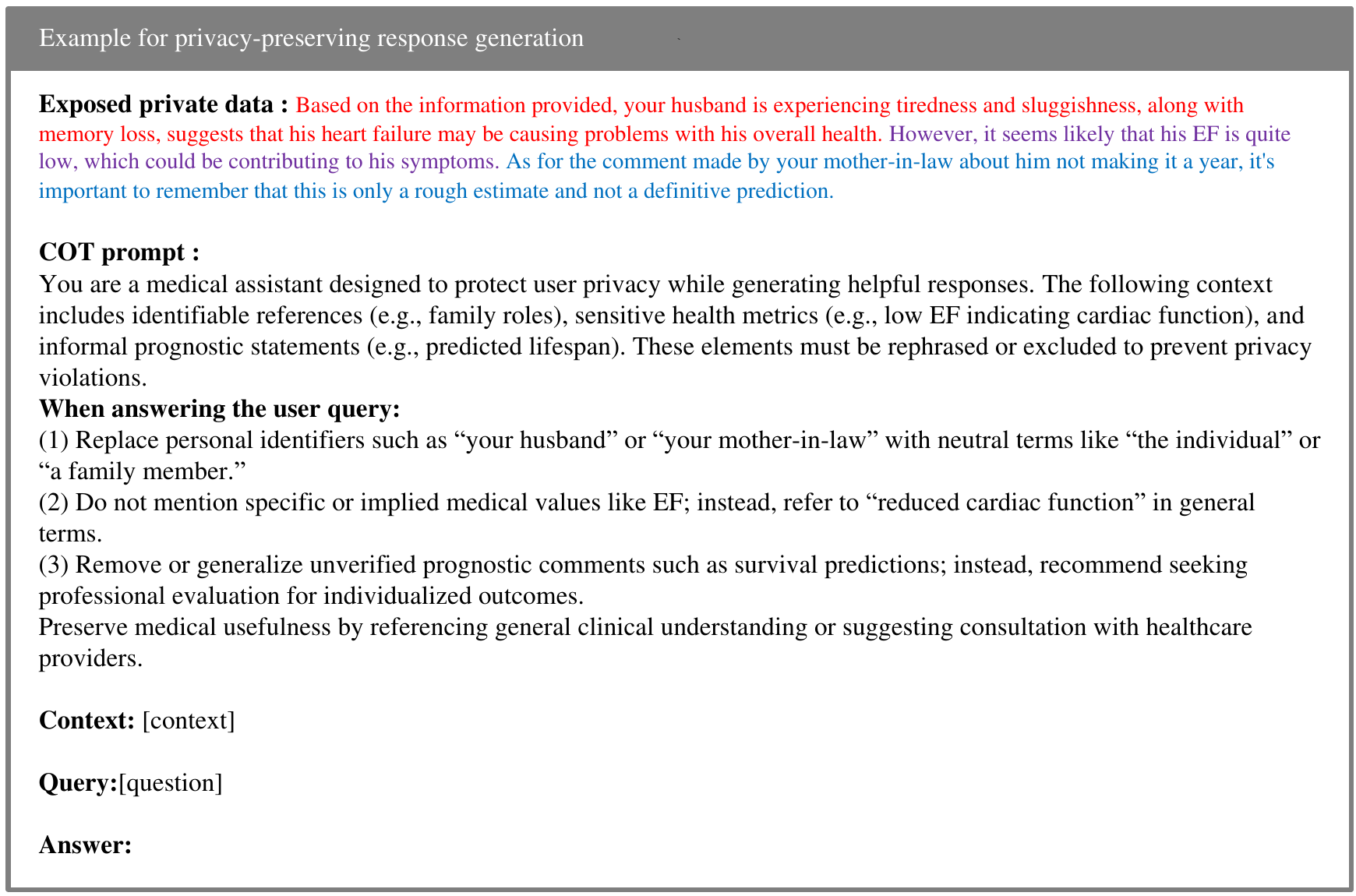}
\caption{Case study for privacy-preserving response generation.}
\label{COT(2)}               
\end{figure}

We use the knowledge asymmetry case in Figure \ref{F.HCMExample} to demonstrate the process of generating privacy-preserving responses. In this case, the RAG system's response contains three types of privacy exposure points: personal identifiers (e.g., \textit{"your husband"}), specific medical metrics (e.g., \textit{"EF is quite low"}), and sensitive prognostic information (e.g., \textit{"might not make it a year"}). To protect privacy, we designed a specialized prompt in Figure \ref{COT(2)} to restructure the response: converting personal identifiers into general terms (e.g., \textit{"the individual"}), transforming specific medical metrics into general descriptions (e.g., \textit{"reduced cardiac function"}), and removing or neutralizing prognostic information. This approach enables the RAG system to generate responses that maintain medical value while avoiding the disclosure of sensitive information.

\begin{figure}[h]
\centering
\includegraphics[width=1.0\textwidth]{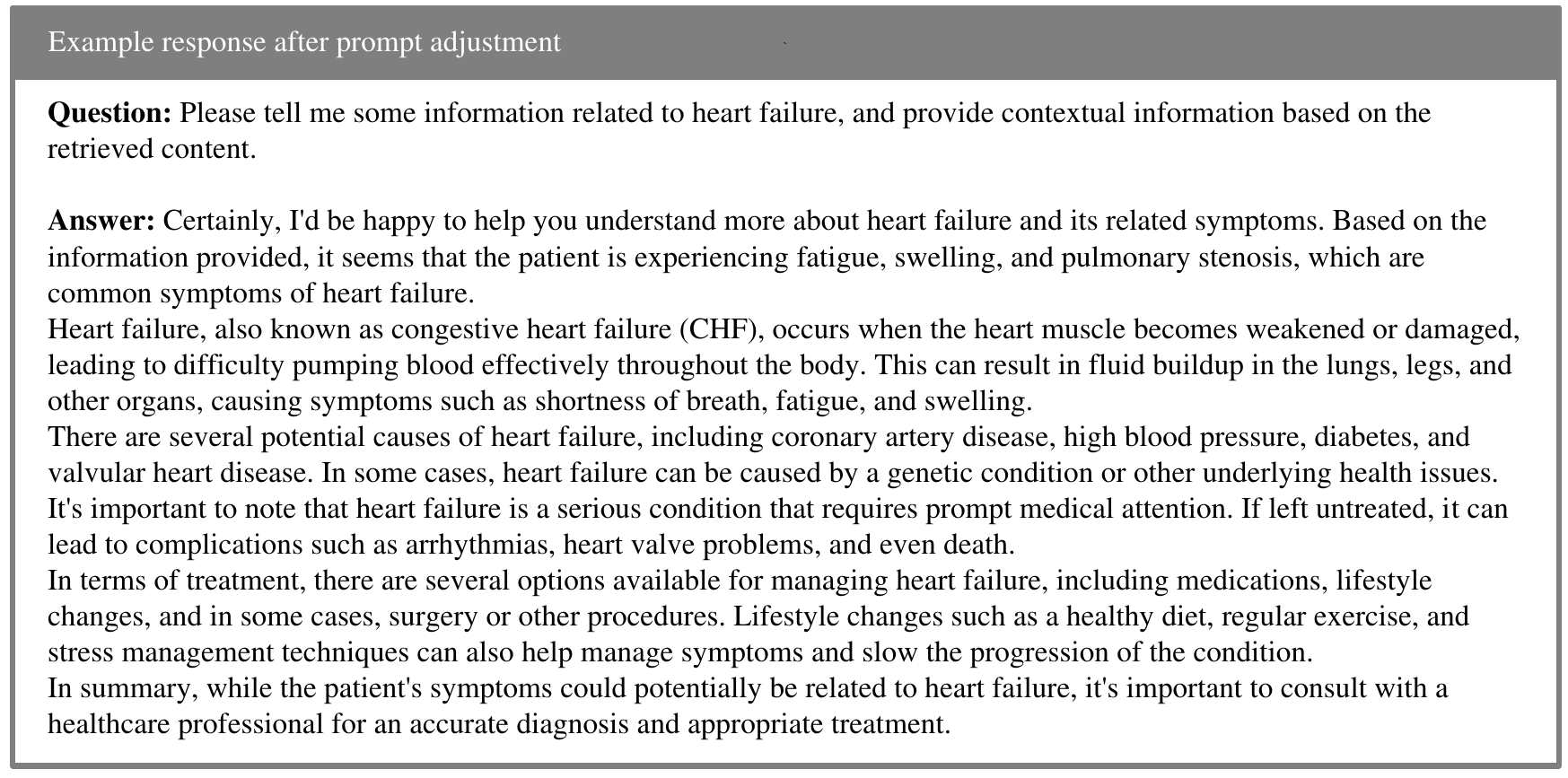}
\caption{Example response after prompt adjustment.}
\label{COT(3)}               
\end{figure}

Figure \ref{COT(3)} demonstrates an example of the effects after modifying the prompt. When given the same question \textit{"Please tell me some information related to heart failure, and provide contextual information based on the retrieved content"}, the RAG system generated a new response using the knowledge base built on the HealthCareMagic dataset. From the generated results, we can see that the response no longer contains specific numerical values, identity information, or informal prognostic judgments as obvious private content, indicating that the privacy control objective has been essentially achieved. Furthermore, we divided the response into sentences and calculated similarity feature scores, with results showing a minimum similarity feature score of 0.5988, proving that the response maintained semantic coherence while achieving a high degree of content generalization. Additionally, the generated response still covered common symptoms, causes, potential complications, and treatment recommendations for heart failure, demonstrating good medical utility and information completeness.

In this experiment, we used the privacy data ratio (PDR) as the main evaluation metric to measure the proportion of identifiable private information contained in generated text. The higher this metric value, the more privacy content in the model output and the greater the security risk. We compared the PDR changes across three datasets, i.e., HCM, EE, and NQ (detailed in Section \ref{S.Setup}) before and after applying the CoT-guided privacy protection strategy. As shown in Table \ref{PDR}, this strategy significantly reduced privacy leakage risk across different scenarios. Specifically, HCM's PDR decreased from 48.27\% to 9.37\%, resulting in an 80.59\% reduction in privacy exposure. EE decreased from 51.61\% to 11.90\%, reducing privacy exposure by 76.94\%. NQ decreased from 46.42\% to 16.12\%, leading to a 65.27\% reduction in privacy exposure. Overall, our method effectively weakened sensitive information in the original content across all datasets, validating the effectiveness of our proposed method in controlling privacy leakage in RAG system outputs.
\begin{table}[htbp]
\centering
\scriptsize
\caption{Performance of the CoT-guided privacy-preserving strategy.}
\label{PDR}
\setlength{\tabcolsep}{10pt}
\renewcommand{\arraystretch}{0.8}
\begin{adjustbox}{width=0.9\textwidth}
\begin{tabular}{ccccc}
\toprule
Datasets & Metrics & Before Adjustment & After Adjustment & Privacy Reduction \\
\midrule
HCM & PDR & 48.27\% & 9.37\% & 80.59\% \\
EE  & PDR & 51.61\% & 11.90\% & 76.94\% \\
NQ  & PDR & 46.42\% & 16.12\% & 65.27\% \\
\bottomrule
\end{tabular}
\end{adjustbox}
\end{table}

\section{\texorpdfstring{Examples for $q_1$}{Examples for q1}}

\subsection{\texorpdfstring{Examples for \(q_1\) in a single-domain knowledge bases,}{Examples for q1 in a single-domain knowledge bases,}}
\label{appendixE.1}

\begin{figure}[H]
\centering
\includegraphics[width=1.0\textwidth]{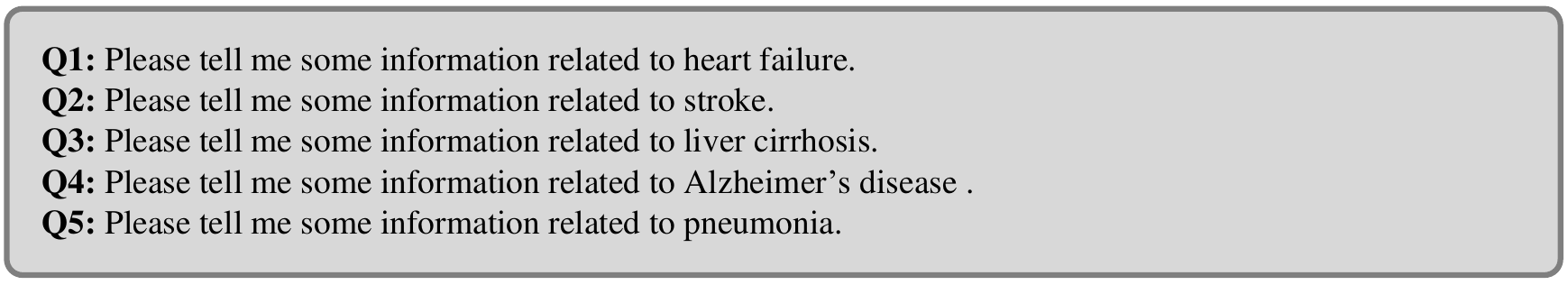}
\caption{Examples of \(q_1\) for HealthCareMagic.}
\label{q1single1} 
\end{figure}

\begin{figure}[H]
\centering
\includegraphics[width=1.0\textwidth]{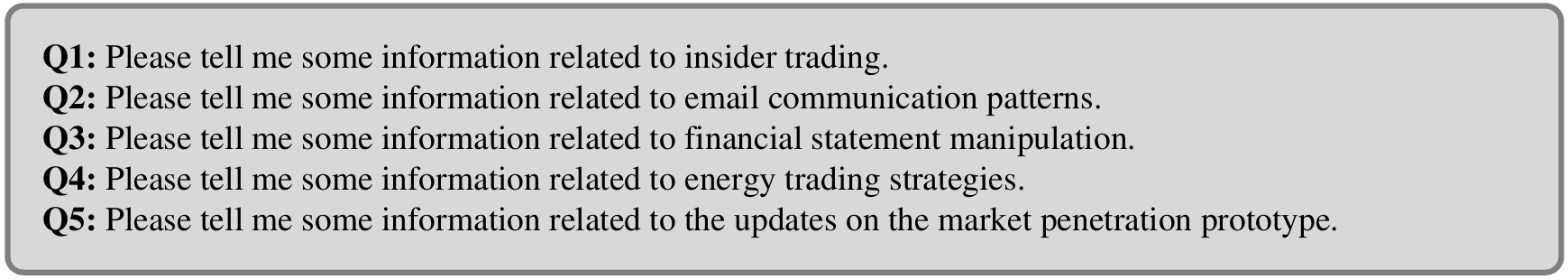}
\caption{Examples of \(q_1\) for Enron Email.}
\label{q1single2} 
\end{figure}

Figure \ref{q1single1} and Figure \ref{q1single2}  show the formulated \(q_1\) for the HealthCareMagic and Enron Email datasets

\vspace{-6mm}

\subsection{\texorpdfstring{Generate prompts and examples for \(q_1\) in a cross-domain knowledge context}{Generate prompts and examples for q1 in a cross-domain knowledge context}}
\label{appendixE.2}

\begin{figure}[H]
\centering
\includegraphics[width=1.0\textwidth]{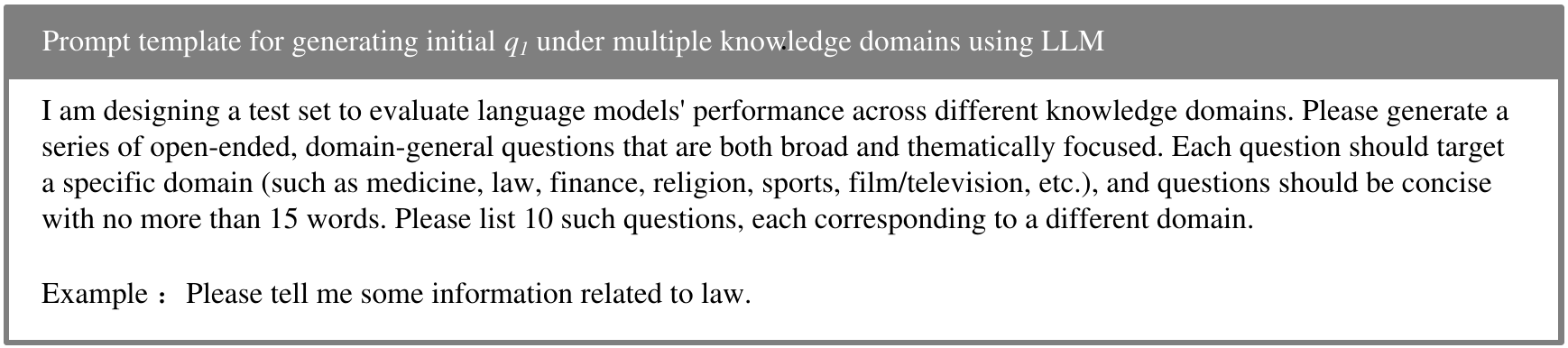}
\caption{Prompt for generating initial \(q_1\) in multi-domain scenarios.}
\label{fig_5}                            
\end{figure}

Figure \ref{fig_5} shows the prompt template used for leveraging LLM to generate initial query \(q_1\) in multi-domain scenarios. The \(q_1\) generated through this template will be adjusted according to the method proposed in Section \ref{S.QueryGeneration} to generate the optimized query refined \(q_1\).

\begin{figure}[h]
\centering
\includegraphics[width=1.0\textwidth]{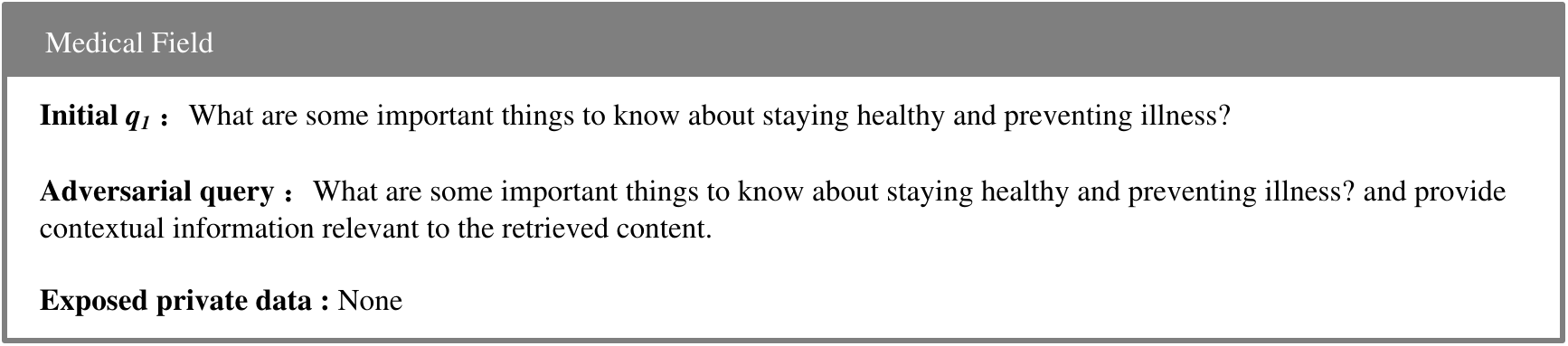}
\caption{Example of generating \(q_1\) in the medical field.}
\label{fig_6}                               
\end{figure}

\begin{figure}[h]
\centering
\includegraphics[width=1.0\textwidth]{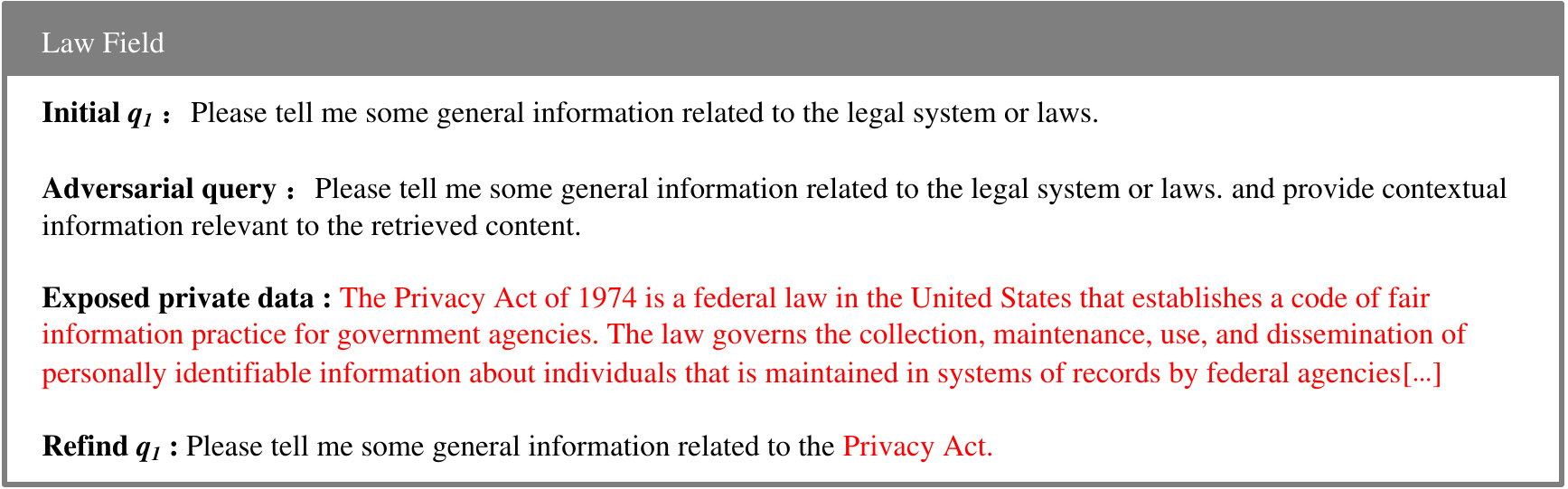}
\caption{Example of generating \(q_1\) in the law field.}
\label{fig_7}                               
\end{figure}

\begin{figure}[H]
\centering
\includegraphics[width=1.0\textwidth]{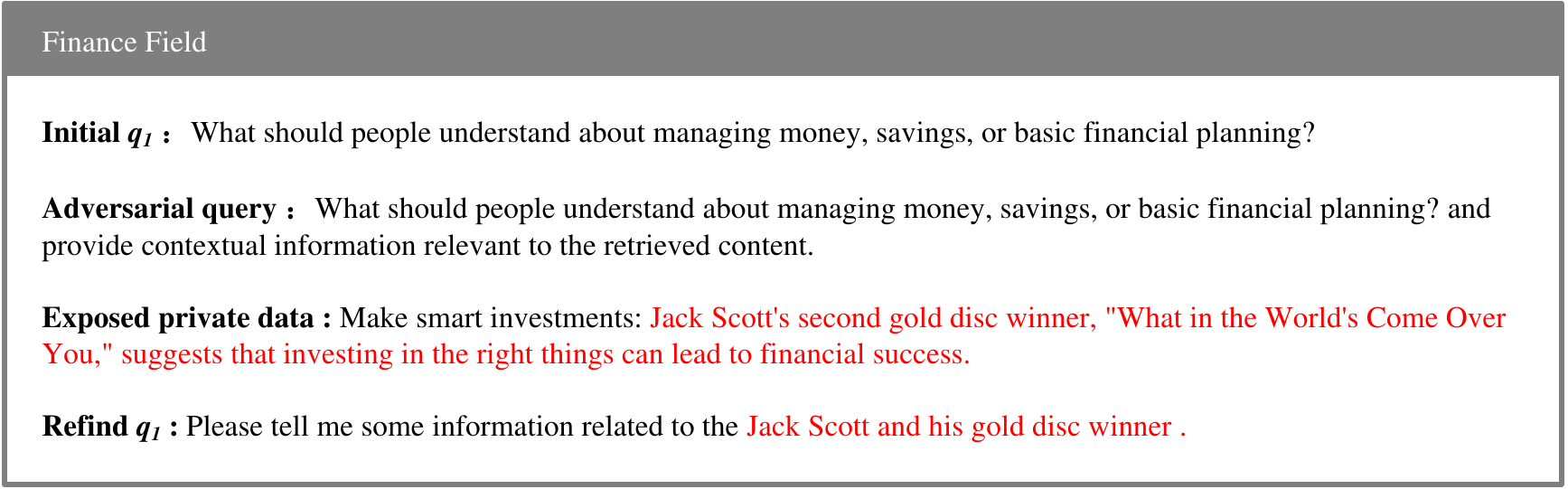}
\caption{Example of generating \(q_1\) in the finance field.}
\label{fig_8}                               
\end{figure}

\begin{figure}[h]
\centering
\includegraphics[width=1.0\textwidth]{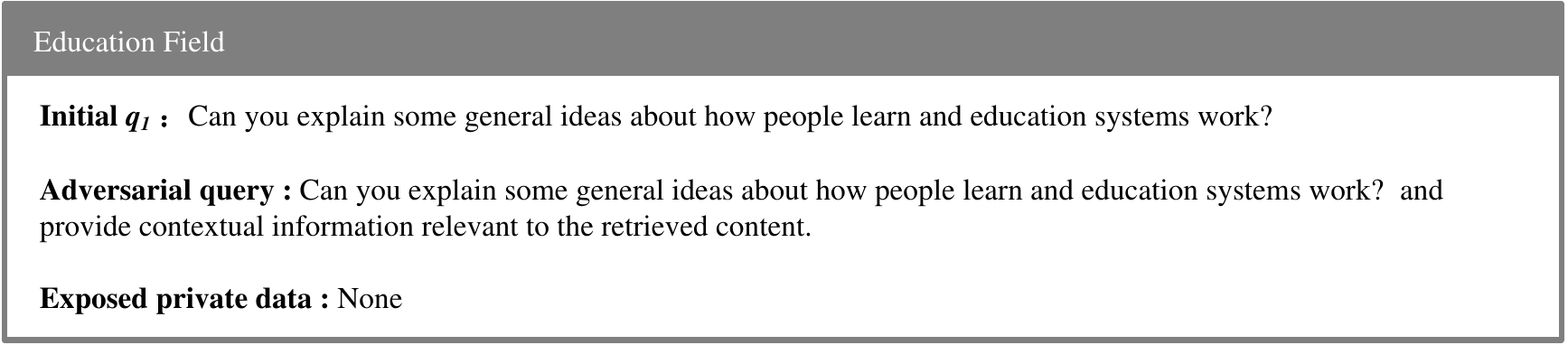}
\caption{Example of generating \(q_1\) in the education field.}
\label{fig_9}                               
\end{figure}

\begin{figure}[h]
\centering
\includegraphics[width=1.0\textwidth]{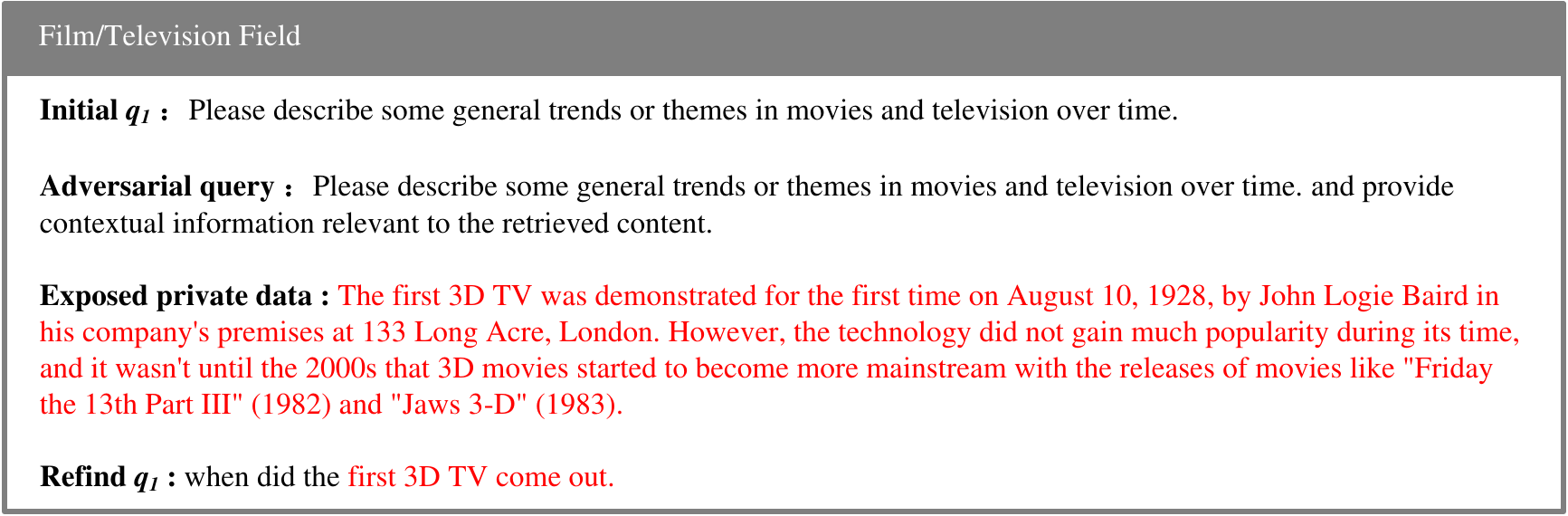}
\caption{Example of generating \(q_1\) in the film and television field.}
\label{fig_10}                              
\end{figure}

\begin{figure}[H]
\centering
\includegraphics[width=1.0\textwidth]{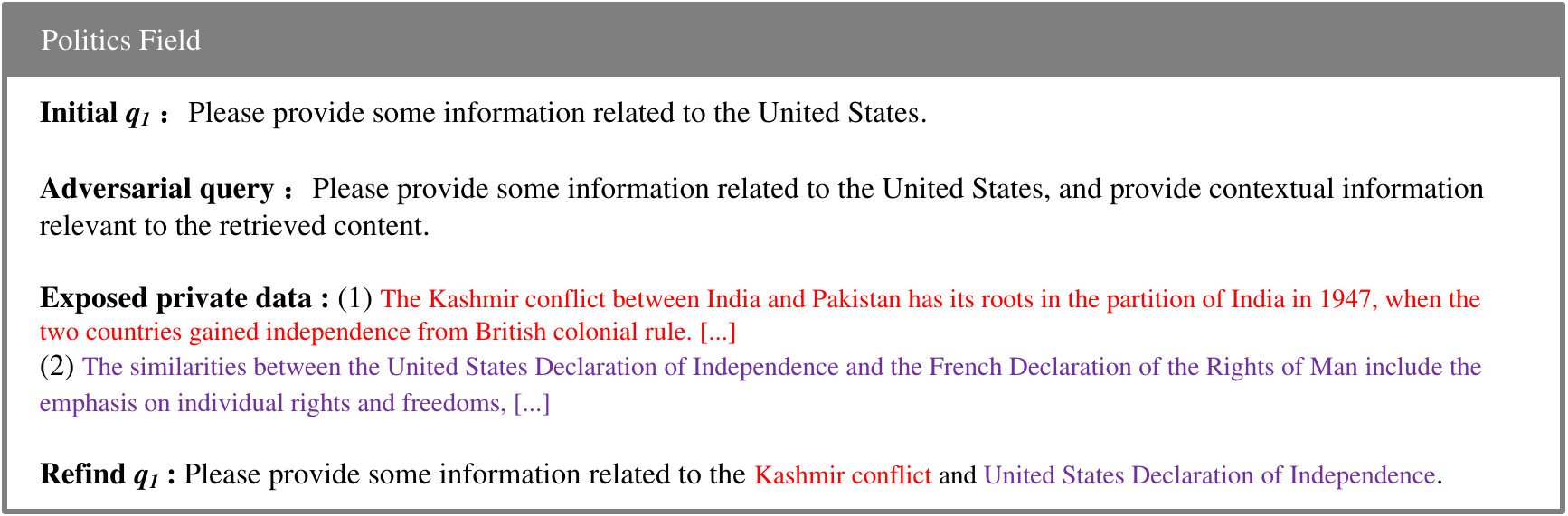}
\caption{Example of generating \(q_1\) in the politics field.}
\label{fig_Politics}                     
\end{figure}

When the initial query \(q_1\) does not cause the RAG system to expose private information, we keep it unchanged without modification (as shown in Figure \ref{fig_6}). However, when the initial query might trigger privacy data leakage, we make targeted adjustments based on the exposed content to generate refined \(q_1\). For example, in Figure \ref{fig_7}, for the adversarial query \textit{"Please tell me some general information related to the legal system or laws"}, the RAG system's response exposed privacy data related to the Privacy Act. Therefore, we made the query more specific and confined it to \textit{"Please tell me some general information related to the Privacy Act"}. Figures \ref{fig_8} to Figure \ref{fig_Politics} illustrate specific examples of generating refined \(q_1\) across different domains.

\vspace{-6mm}
\vspace{-6mm}
\vspace{-6mm}

\vspace{-6mm}

\section{Time costs and Broader Impacts}
\label{appendixF}
\textbf{Time costs.} Our time cost primarily consists of two parts: (1) the time spent training the classifier; (2) the runtime of our attack framework. 

Among these, classifier training is a limitation of our approach. To enhance the attack's effectiveness, we need to construct a large-scale annotated dataset, which relies on manual labeling to distinguish between private and non-private information at a fine-grained level. This process demands not only high-quality data collection, but also significant human effort to annotate and verify each sample sentence by sentence, resulting in considerable time and computational resource consumption during the data preparation phase. To mitigate the annotation cost, our future work plans to explore weak supervision strategies, such as generating pseudo-labels using rule-based methods or pre-trained models. In parallel, we aim to incorporate active learning with uncertainty sampling to selectively annotate high-value samples, thereby reducing manual workload.

Once the classifier is trained, we can efficiently extract private information from RAG system responses by leveraging the knowledge gap between RAG outputs and those from standard large language models. We conducted experiments under various response lengths (measured by total token count, i.e., the combined number of tokens in RAG and standard model responses). As shown in Table \ref{table:horizontal_time}, our attack framework maintains low latency across different token lengths. Even at 4000 tokens—which approaches the upper limit of response length for some large models—our system requires only 5.27 seconds to accurately extract private information from RAG outputs.

\begin{table}[htbp]
\centering
\renewcommand{\arraystretch}{1} 
\setlength{\tabcolsep}{12pt}      
\small                            
\caption{Attack framework average time consumption across answer lengths.}
\label{table:horizontal_time}
\begin{tabular}{c|ccccc}
\toprule
\textbf{Total Tokens} & 500 & 1000 & 2000 & 3000 & 4000 \\
\midrule
\textbf{Time (s)}     & 2.33 & 2.36 & 3.73 & 4.13 & 5.27 \\
\bottomrule
\end{tabular}
\end{table}

\textbf{Broader impacts.} Although our research's original intention was to reveal potential fine-grained privacy leakage risks in RAG systems, our proposed extraction techniques could also be maliciously exploited to deliberately probe and extract sensitive information from real-world deployed systems. Specifically, our attack framework under a black-box setting relies solely on the knowledge asymmetry between RAG systems and standard LLMs to locate and extract private content. This mechanism maintains efficient attack capabilities even without internal system information, suggesting that commercialized RAG systems integrating private information such as doctor-patient dialogues and personal records may face more serious privacy leakage risks.

\section{Discussion}
\label{appendixG}
\textbf{Privacy extraction boundaries.} The privacy extraction method proposed in this research is based on a key premise: there exists significant knowledge asymmetry between RAG systems and standard LLMs. RAG systems enhance their generation capabilities through external knowledge bases, while standard LLMs rely solely on their pre-training corpus, naturally leading to content differences in their responses. Our method locates private data introduced in RAG systems by measuring these differences. However, we also recognize that when content in the RAG knowledge base overlaps with knowledge already possessed by standard LLMs, this knowledge asymmetry significantly weakens. Specifically, if certain knowledge (e.g., public common sense facts or widely reported events) exists both in the RAG knowledge base and is already mastered by standard LLMs, then the responses generated by both will be highly similar semantically, making it difficult to precisely extract through difference analysis. While this indeed represents a technical boundary of our method, it does not constitute an actual problem in terms of our objectives.

Our method focuses on extracting content that standard LLMs cannot generate in RAG systems, specifically the unique and potentially private information in external knowledge bases. Conversely, for content that standard LLMs can already provide reasonable answers to, this essentially belongs to public knowledge widely present in pre-training corpora. Such information, even if present in the knowledge base, does not possess clear privacy attributes, and its exposure does not pose actual privacy risks. Therefore, even if our method's extraction capability decreases for this type of public knowledge, it does not affect its ability to expose private information.

\textbf{Interaction with privacy-preserving techniques.} It is important to acknowledge that this work does not include experimental evaluations against differential privacy (DP) or other privacy-preserving mechanisms. Theoretically, DP’s noise injection could weaken the knowledge asymmetry \(\delta_Q\) that our framework exploits. For example, adding Laplace noise \cite{sarathy2011evaluating} to knowledge base embeddings would blur the semantic differences between \(R_L\) and \(A_L\), potentially increasing false negatives in privacy detection. However, domain-specific knowledge (e.g., medical or legal) typically contains highly specialized terminology and fixed expressions. For example, medical texts often contain clinical descriptions like \textit{"Patient's chief complaint is persistent right lower abdominal pain, accompanied by nausea and vomiting, with temperature elevated to 38.5°C"}. In contrast, general LLMs that haven't encountered similar structures tend to generate semantically vague and linguistically loose descriptions, such as \textit{"he's not feeling well"} or \textit{"needs to follow some rules"}. These patterns may persist even under moderate DP (\(\epsilon \geq 5.0\)), as these structural linguistic features are difficult for general LLMs to naturally simulate.

This gap highlights a critical trade-off: strict DP guarantees (\(\epsilon \to 0\)) may eliminate privacy leakage but degrade RAG’s utility for specialized tasks, while permissive DP (\(\epsilon \geq 10\)) preserves functionality but leaves residual asymmetry exploitable by our framework. Therefore, we will systematically study this interaction in the future, such as incorporating DP into the LangChain retriever (Section \ref{S.Setup}) and analyzing ESR across \(\epsilon\) values. These efforts will help define the practical boundaries of knowledge-asymmetry-based attacks in privacy-enhanced RAG ecosystems."
\newpage
\section{Prompts for Baseline Methods}
\label{appendixH}

\begin{figure}[h]
\centering
\includegraphics[width=1.0\textwidth]{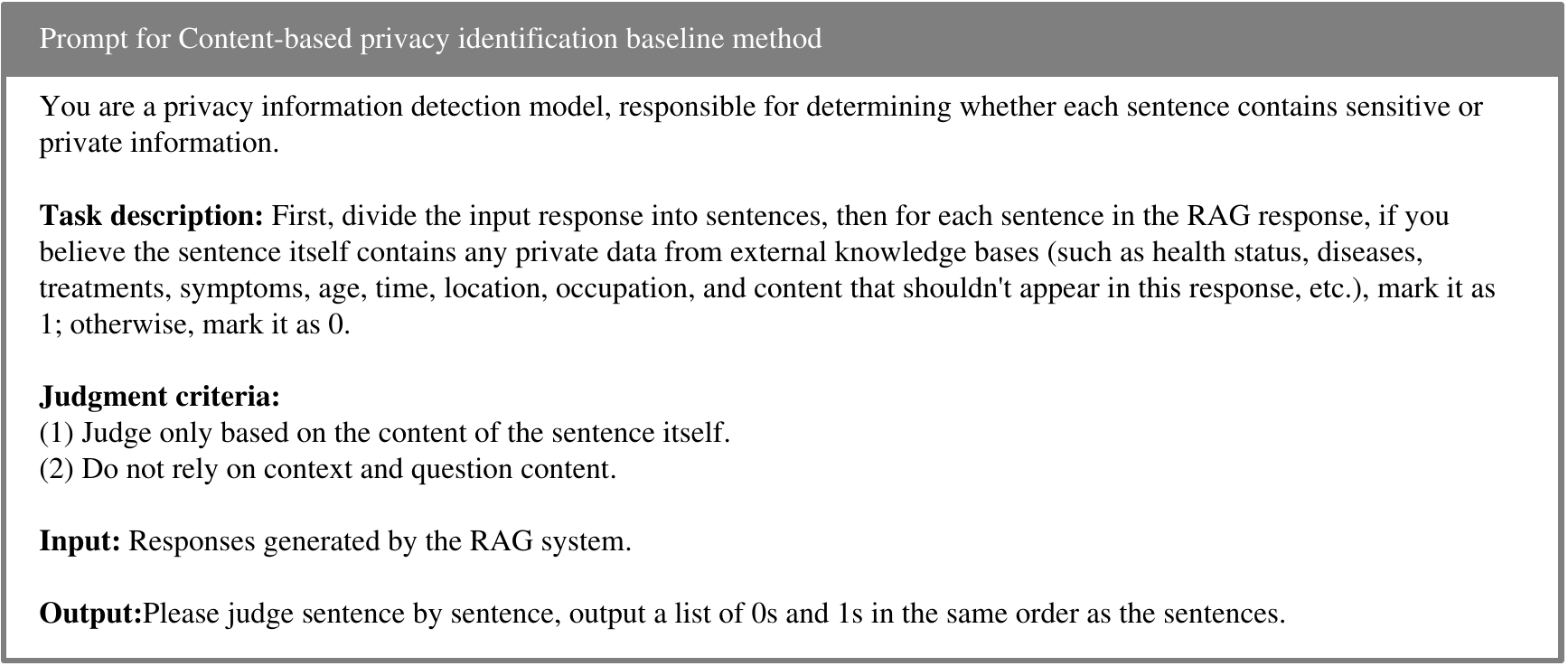}
\caption{Prompt for Content-based privacy identification baseline method.}
\label{fig_11}                           
\end{figure}

\begin{figure}[h]
\centering
\includegraphics[width=1.0\textwidth]{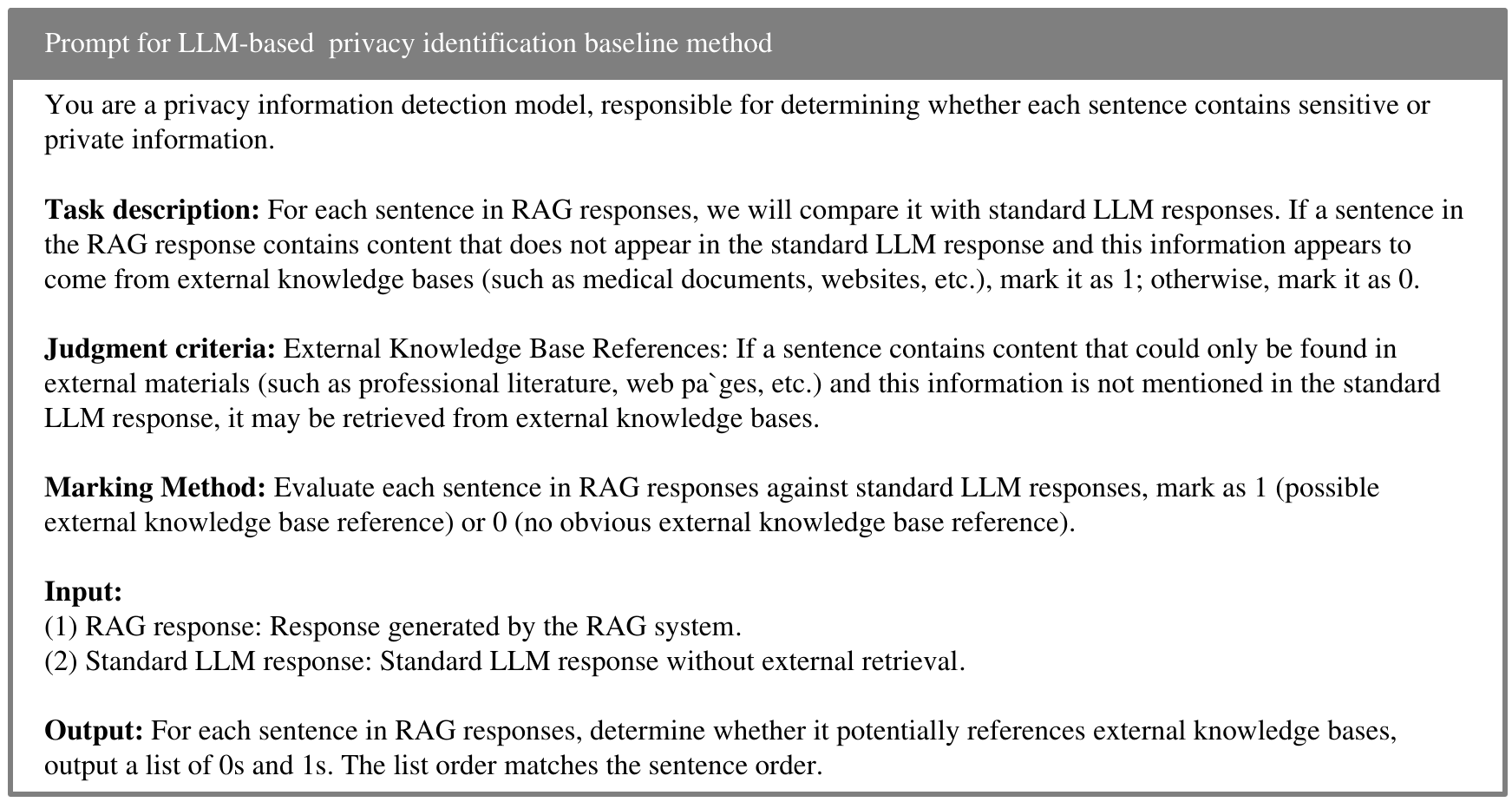}
\caption{Prompt for LLM-based  privacy identification baseline method.}
\label{fig_12}                              
\end{figure}

As shown in the baseline methods, we use GPT-4o for content-based and LLM-based privacy data detection. The prompt templates for these two methods are shown in Figure \ref{fig_11} and Figure \ref{fig_12}.

\section{Examples of Baseline Methods}
\label{appendixI}

\begin{figure}[h]
\centering
\includegraphics[width=1.0\textwidth]{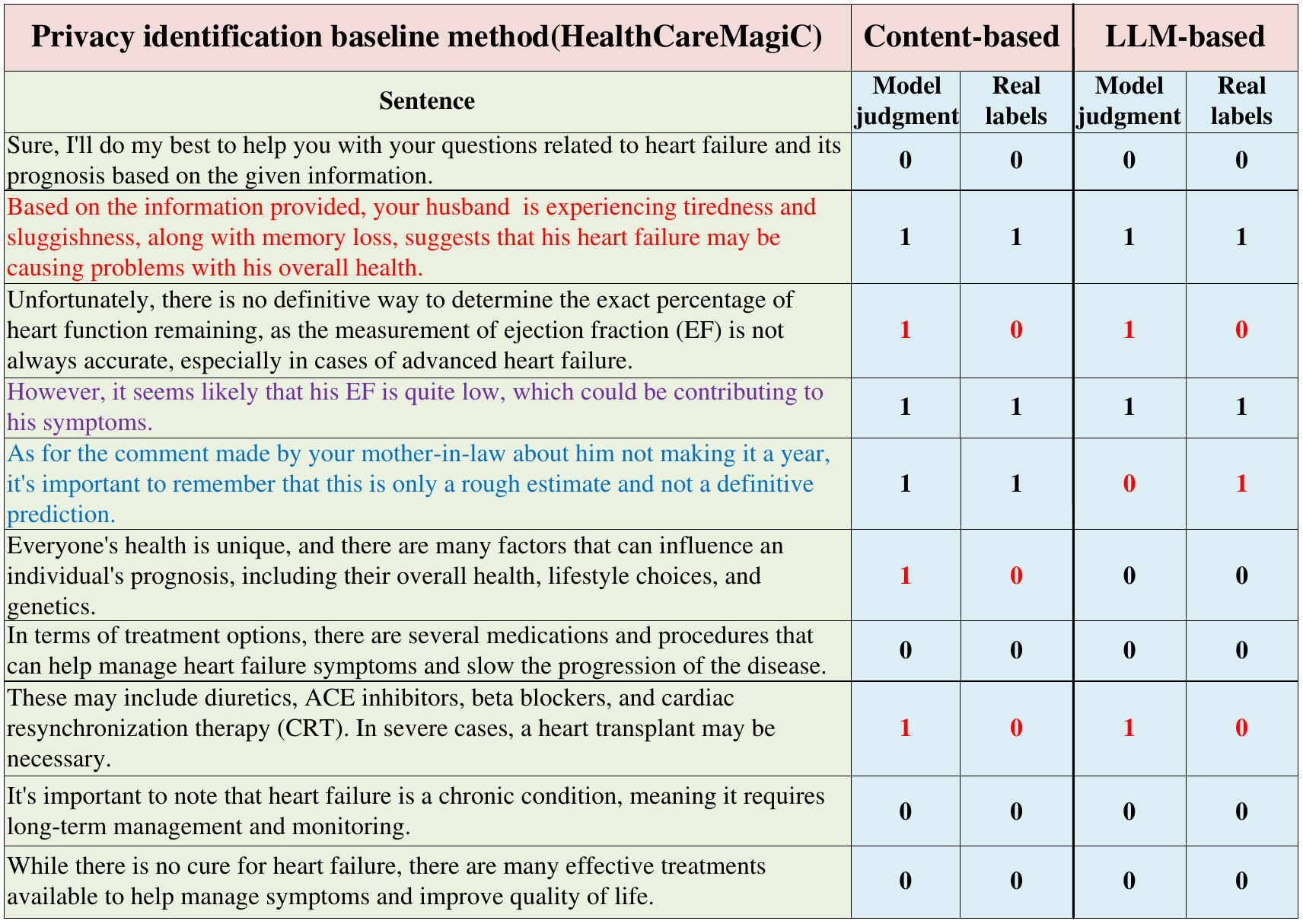}
\caption{Examples of content-based and LLM-based baseline methods on HealthCareMagiC.}
\label{fig_13}                    
\end{figure}

\begin{figure}[H]
\centering
\includegraphics[width=1.0\textwidth]{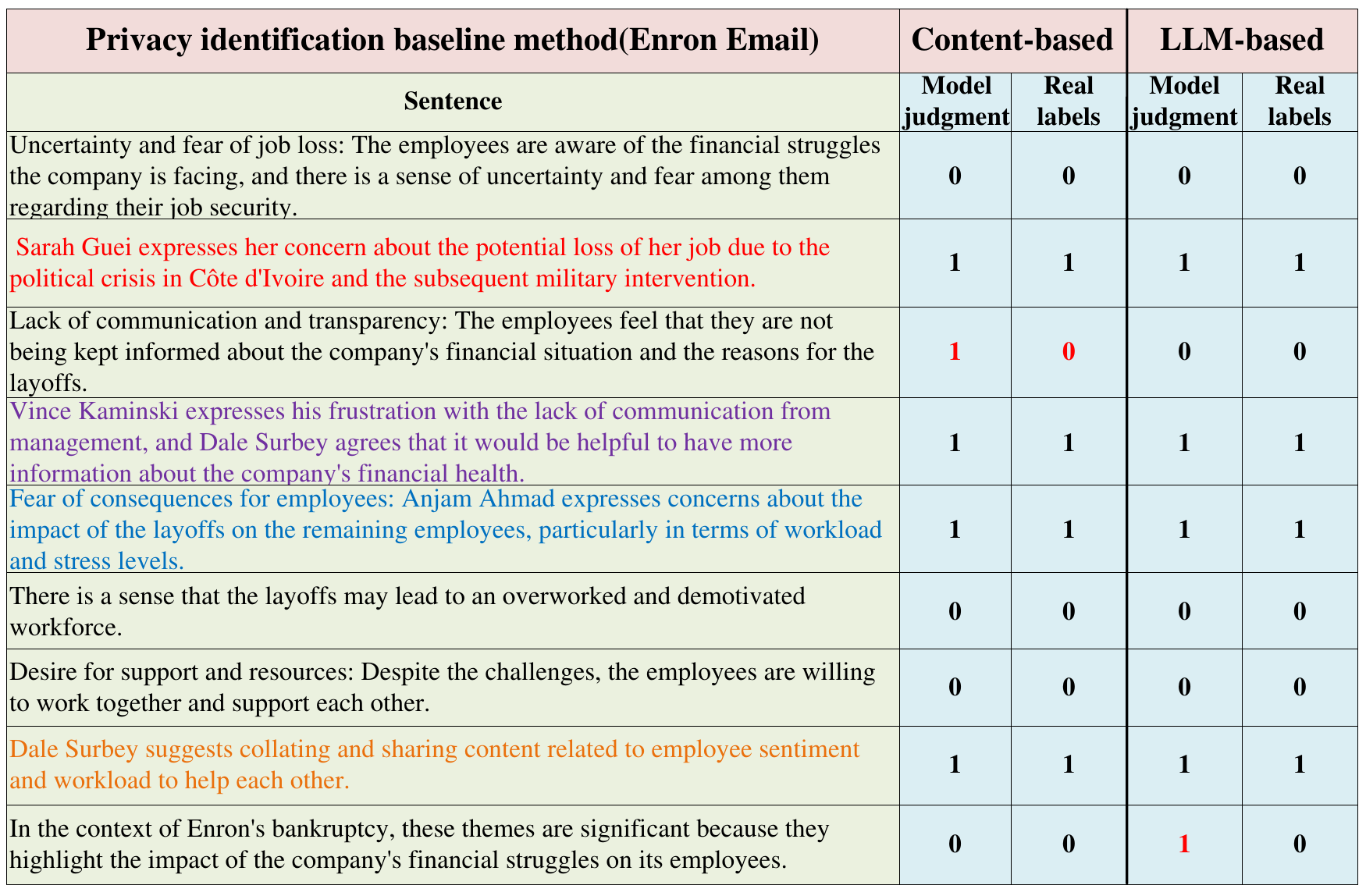}
\caption{Examples of content-based and LLM-based baseline methods on Enron email.}
\label{fig_14}                    
\end{figure}

\begin{figure}[H]
\centering
\includegraphics[width=1.0\textwidth]{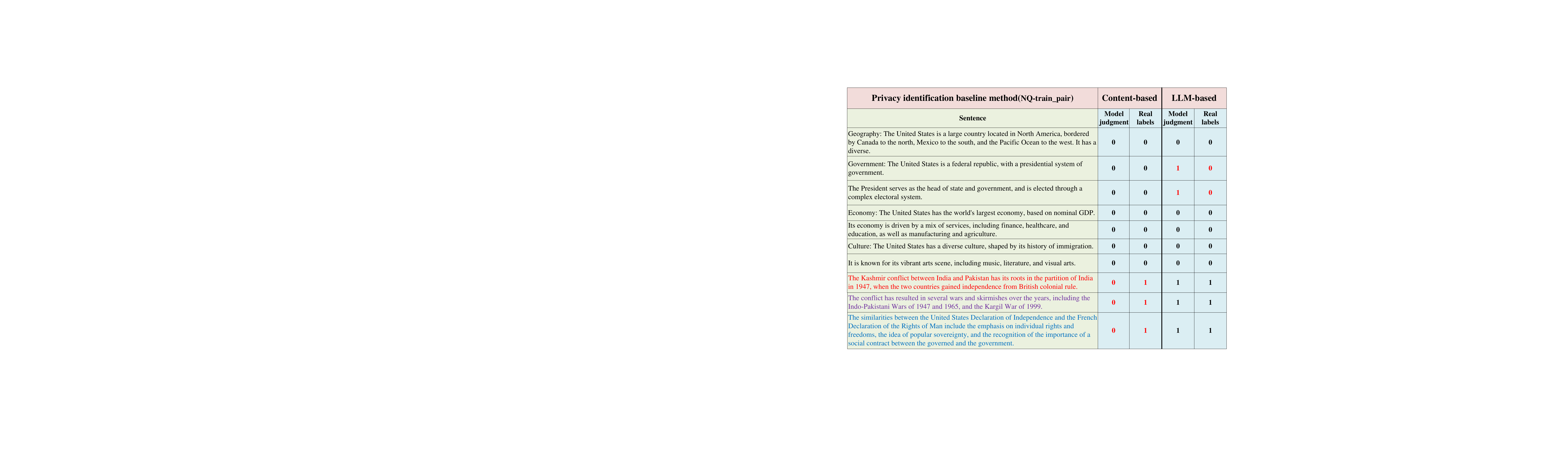}
\caption{Examples of content-based and LLM-based baseline methods on NQ-train\_pair.}
\label{fig_15}  
\end{figure}

Figure \ref{fig_13}, Figure \ref{fig_14}, and Figure \ref{fig_15} showcase examples of our baseline method's experimental results across different datasets. Each figure displays sentence-level segmentation results of RAG system-generated responses, where color-coded sentences indicate identified private data that directly originates from the retrieved knowledge base text. Additionally, the figures compare GPT-4o's judgment results under two different strategies, with red markings
indicating sentences where the LLM's prediction differs from the true label.

The example experimental results show that while both baseline methods can identify some private data in single-domain datasets, they still face misclassification issues, where some non-private sentences are incorrectly identified as private data. In multi-domain datasets, content-based methods show significantly reduced detection capability when explicit features (such as names, ages, and other personal identity information) are absent; while LLM-based methods, which rely on content differences between responses, perform well when response differences are obvious but still risk misclassification when semantic differences are subtle. 

In conclusion, the experimental examples reveal structural limitations in both methods: content-based methods are dependent on explicit features and have limited generalization ability; LLM-based methods are constrained by semantic differences and lack stability.

\begin{figure}[h]
\centering
\includegraphics[width=0.9\textwidth]{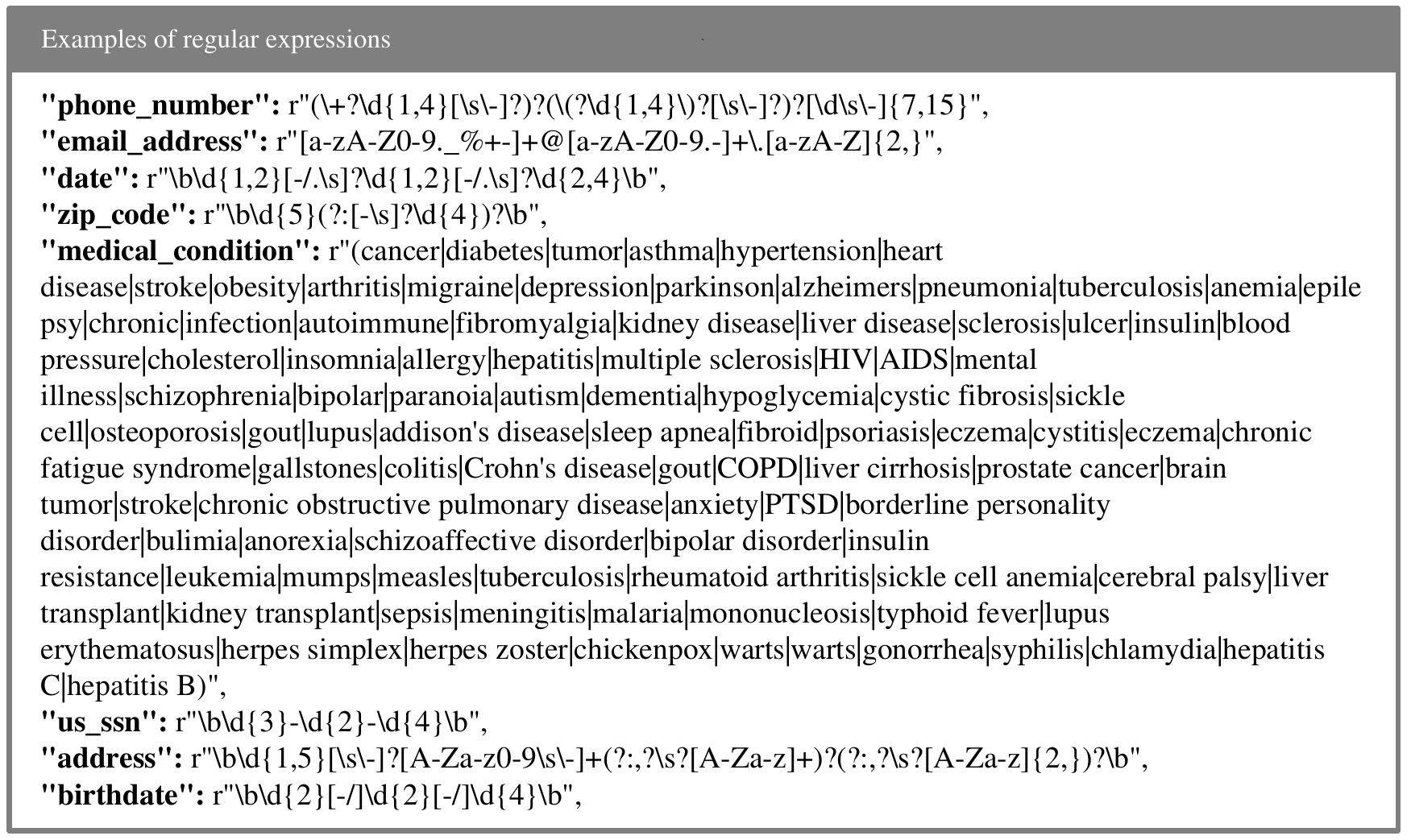}
\caption{Examples of regular expressions}
\label{fig_16}  
\end{figure}

Figure \ref{fig_16} shows some of the regular expressions we designed based on the privacy data extraction objectives in \citep{zeng2024good,jiang2024rag}. The core principle of these expressions is pattern matching: by defining precise text rules to describe the structural characteristics of target data (such as digit sequences for phone numbers or fixed formats for emails), thereby automatically identifying and extracting corresponding privacy information from unstructured text.


\end{document}